\documentclass{article}

\usepackage{amssymb,amsfonts,amsmath,stmaryrd}
\usepackage{cite,enumerate,float,indentfirst}
\usepackage{rotating}

\def\be{\begin{eqnarray}}
\def\ee{\end{eqnarray}}
\def\nn{\nonumber}

\def\tr{{\rm tr}\,}

\hoffset= - 1.0in         

\begin{document}

\hfill MIPT/TH-03/20

\hfill ITEP/TH-03/20

\hfill IITP/TH-03/20

\bigskip

\centerline{\Large{ KNTZ trick from arborescent calculus
and the structure of differential expansion
}}

\bigskip

\centerline{{\bf A.Morozov}}

\bigskip

\centerline{\it MIPT, ITEP \& IITP, Moscow, Russia}

\bigskip

\centerline{ABSTRACT}

\bigskip

{\footnotesize
The recently suggested KNTZ trick completed the lasting search for exclusive
Racah matrices $\bar S$ and $S$ for all rectangular representations
and has a potential to help in the non-rectangular case as well.
This was the last lacking insight about the structure of differential expansion
of (rectangularly-)colored knot polynomials for twist knots -- and the resulting success
is a spectacular achievement of modern knot theory in a classical field
of representation theory, which was originally thought to be a tool for
knot calculus but instead appeared to be its direct beneficiary.
In this paper we explain that the KNTZ ansatz is  actually a suggestion to
convert the arborescent evolution matrix $\bar S\bar T^2\bar S$ into triangular form
${\cal B}$  and demonstrate how this works and what is the form of the old puzzles
and miracles of the differential expansions from this perspective.
The main new fully result is the conjecture for the triangular matrix ${\cal B}$ in the
case of non-rectangular representation $[3,1]$.
This paper does not simplify any calculations, but highlights the remaining problems,
which one needs to overcome in order to {\it prove} that things really work.
We believe that this discussion is also useful for further steps towards
non-rectangular case and the related search of the gauge-invariant
arborescent vertices.
As an example we formulate a puzzling, still experimentally supported conjecture,
that the study of twist knots only
is sufficient to describe the shape of the differential expansion for all knots.

}

\bigskip

\bigskip

\section{Introduction}

This paper is about the existence and implications of the {\it differential expansion}
\cite{DGR,IMMMfe,evo,diffarth}
for colored knot polynomials \cite{knotpols,Wit,ind,RT}
\be
H_R^{\cal K}(A,q) = \sum_{X\in {\cal R}_R} Z_R^X(A,q) F_X^{\cal K}(A,q)
\ee
which separate their dependencies on knots ${\cal K}$ and representations $R$.
Since the use of this expansion for evaluation of exclusive Racah
matrices \cite{Racah} and for arborescent calculus \cite{arbor} along the lines,
originally suggested in \cite{M16}, was reviewed  very recently in \cite{KNTZ,M19,M19nr},
we avoid doing this once again
and directly proceed to considerations, outlined in the abstract of this paper.
They touch three main issues.

The {\bf first} is the KNTZ claim \cite{KNTZ}
that the switch from a diagonal evolution matrix $\bar T^2$ to triangular
${\cal B}$, though looks like a complication,
actually reveals the hidden structure of the differential expansion for twist knots
and somehow trades the sophisticated Racah matrix $\bar S$ for a much simpler and
universal (representation-independent) ${\cal B}$.
Following \cite{M19pentad},
we suggest that the reason for this can be that the actual evolution matrix
was not the simple diagonal $\bar T^2$,
but rather a sophisticated symmetric $\bar S\bar T^2\bar S$,
and then the switch to triangular ${\cal B}$ is indeed a simplification.
In this approach the crucial role is played by the switching matrix $U$,
and the central phenomenon is a drastic simplicity of the first line
in a peculiar matrix $U\bar T^2U^{-1}{\cal B}^{-1}$:
for rectangular representations
its entries are just products of the differentials.
Better understanding of the phenomenon can help to explain what are
the linear combinations of those, which emerge
for  non-rectangular representations.

The {\bf second} issue is to explain the main differences in non-rectangular case.
The crucial one is emergency of multiplicities -- and we consider explicit
examples of representations $[2,1]$ and $[3,1]$ to demonstrate what is their role.
In twist knot polynomials the multiplicities can be largely ignored, in the sense
that there exists a version of triangular ${\cal B}$ without multiplicities --
as already argued in \cite{M19nr}.
However, already to tame the $Z$-factors and, further, to find the Racah matrices \cite{Racah},
the full ${\cal B}$ in the full multiplicity space are needed.
An amusing fact is that still some pieces of ${\cal B}$ and $\bar S$ decouple.
In numbers this looks as follows: for representation $[2,1]$ the full matrices
${\cal B}$ and $\bar S$ are
$10\times 10$, but they split into the relevant $8\times 8$ blocks ${\cal B}^{rel}$,
${\bar S}^{rel}$ and irrelevant $2\times 2$ blocks,
while to describe knots it is enough to look at the reduced $6\times 6$ matrix ${\cal B}^{\rm red}$.
We also list multiplicities for more complicated representations,
but building up the corresponding matrices ${\cal B}$ etc requires new techniques
and will be discussed elsewhere.

The {\bf third} issue is the structure of the differential expansion for arbitrary knots,
where nothing like matrix ${\cal B}$ exists.
Still one can assume that the number of items and the corresponding $Z$-factors are
always the same as for the twist knots -- and we explicitly formulate this conjecture
at the end of this paper.

\bigskip

Unifying the three topics is the question of how much of exclusive Racah matrices $\bar S$
and $S$ is contained in ${\cal B}$.
The point is that ${\cal B}$ looks much simpler, moreover it is explicitly known
from \cite{KNTZ,M19}
for all rectangular representations and even for their Macdonald deformations,
which lead to hyper- and, presumably, superpolynomials
\cite{GSV,DGR,DMMSS,Okshift,Okmac,GGS,NawOb,Anokhevo}.
Even more than that, it looks like reduced matrix ${\cal B}^{\rm red}$ can be also found
for non-rectangular representations.
Therefore it is important to understand, if and how one can reconstruct $\bar S$ from
the knowledge of ${\cal B}$ -- and what substitutes ${\cal B}$ or
what survives from the entire structure (if anything) for non-twist knots,
at least arborescent.
There is still no satisfactory answer for all these questions,
and in this paper we review various attempts and concrete examples
for small representations -- which involve additional matrices $U$ and ${\cal E}$,
the two extra members of the pentad \cite{M19pentad},
lying somewhere in between ${\cal B}$ and $\bar S$.
The hope is that they are easier to guess than $\bar S$, but contain more information
than ${\cal B}$.

\section{$U$-matrix and its properties}

Arborescent formula \cite{arbor} for the normalized 
HOMFLY-PT polynomial of the twist knot
\be
H_R^{\text{twist}_m} =
d_R \cdot \Big(\bar S_R \bar T^{2} \bar S_R \bar T^{2m} \bar S_R\Big)_{\emptyset\emptyset}
\label{arborHOMFLY}
\ee
with symmetric and orthogonal Racah matrix $\bar S_R$, $\bar S_R^2=I$,
can be identically transformed into
\be
H_R^{\text{twist}_m}
= d_R \cdot \Big(\bar S_R \bar T^{2}    \
(\bar S_R \bar T^{2} \bar S_R)^{m }\Big)_{\emptyset\emptyset}
= d_R \cdot \Big(\bar S_R \bar T^{2} \bar S_R \bar T^{-2} \bar S_R\
(\bar S_R \bar T^{2} \bar S_R)^{m+1}\Big)_{\emptyset\emptyset}
\ee
and then rewritten in the KNTZ  form \cite{KNTZ,M19,M19nr,M19pentad}
of the differential expansion \cite{DGR,IMMMfe,evo,diffarth},\cite{M16},\cite{pretzel}--\!\cite{M333}:
\be
H_R^{\text{twist}_m} = \sum_{X }
\overbrace{d_R \cdot \Big(\bar S_R \bar T^{2}
\underbrace{\bar S_R \bar T^{-2} \bar S_R U_R^{-1}}_{U_R^{-1}{\cal B}^{-1}}
\Big)_{\emptyset X}}^{Z_R^X} \cdot \overbrace{\sum_Y
\Big(\boxed{\underbrace{U_R\bar S_R \bar T^{2} \bar S_R U_R^{-1}}_{{\cal B}}}\Big)^{m+1}_{XY}
\cdot U_{Y\emptyset}}^{F_X^{(m)}}
= \sum_{X} Z_R^X \cdot \overbrace{\sum_Y \left({\cal B}^{m+1}\right)_{XY}}^{F_X^{(m)}}
\label{arbortoKNTZ}
\ee
This is achieved
by additional insertion of the unity decompositions $I=U_R^{-1}U_R$ and $I=S_R U_R^{-1}U_RS_R$
between all the squares $\bar T^2$.
The last transition in (\ref{arbortoKNTZ}) requires that
the auxiliary matrix $U_R$, { which is one of the main heroes of the present paper},
has unities everywhere in the first column,
\be
U_{Y\emptyset} =1
\label{UY0}
\ee
while its other elements are adjusted to make the KNTZ matrix \cite{KNTZ}
\be
\boxed{
{\cal B} := U_R\bar S_R \bar T^{2} \bar S_R U_R^{-1}
}
\label{BviaU}
\ee
{\it triangular} and satisfying the constraints
\be
\sum_{Y} {\cal B}_{XY} = \delta_{X,\emptyset} \ \ \ \ \ \ \forall X
\label{sumB}
\ee
\vspace{-0.5cm}
and
\be
\sum_Y ({\cal B}^2)_{XY} = \sum_{Y,Z} {\cal B}_{XZ}{\cal B}_{ZY}
\ \stackrel{(\ref{sumB})}{=} \ {\cal B}_{X\emptyset}
\ \ \ \ \ \ \forall X
\ee
with ${\cal B}_{XX}=\Lambda_X$
and ${\cal B}_{X\emptyset}=  \Lambda_X'$
being the known monomials, see (\ref{Lambdas}) below.
Remarkably, after $U_R$ is adjusted to convert symmetric $\bar S_R\bar T \bar S_R$
into triangular ${\cal B}$, the matrix elements
\be
\boxed{
Z_R^X :=  d_R \cdot
\Big(\bar S_R \bar T^{2} \bar S_R \bar T^{-2} \bar S_R U_R^{-1} \Big)_{\emptyset X}
}
\label{ZviaU0}
\ee
appear to be factorized for all {\bf rectangular representations} $R$
(for non-rectangular $R$ they are factorized in a special basis for $X$,
otherwise they are sums of several factorized expressions, see \cite{Mnonrect})
and reproduce the hook formulas for the Z-factors in
the differential expansions, in particular
\be
Z_R^\emptyset = d_R \cdot \Big(\bar S_R \bar T^{2} \bar S_R \bar T^{-2}
\bar S_R U_R^{-1} \Big)_{\emptyset \emptyset}\ \stackrel{?}{=}\ 1 \ \ \ \ \forall R
\ee
Question marks here and below express  not a doubt in the validity of the statements,
but the lack of explanation/proof why they are always true.
One more impressive fact is the factorization property,
which was the starting observation of \cite{M16}:
\be
H_R^{\text{double braid}_{m,n}} \stackrel{?}{=}
d_R \cdot \Big(\bar S_R \bar T^{2n} \bar S_R \bar T^{2m} \bar S_R\Big)_{\emptyset\emptyset}
= \ \sum_X Z_R^X \cdot \frac{F_X^{(m)}\cdot F_X^{(n)}}{F_X^{(1)}}
\ee
It is now equivalent to a mysterious identity
\be
\Big(\bar S_R \bar T^{2n} \bar S_R \bar T^{2m} \bar S_R\Big)_{\emptyset\emptyset}
\stackrel{?}{=}\
\sum_X  \Big(\bar S_R \bar T^{2} \bar S_R \bar T^{-2} \bar S_R U_R^{-1} \Big)_{\emptyset X}
\frac{\Big(U_R\bar S_R \bar T^{2(m+1)} \bar S_R \Big)_{X\emptyset}
\Big(U_R\bar S_R \bar T^{2(n+1)} \bar S_R \Big)_{X\emptyset}}
{ \underbrace{\Big(U_R\bar S_R \bar T^{2} \bar S_R U_R^{-1}\Big)_{X\emptyset}}_{\Lambda'_X}}
\label{mystid}
\ee
and can serve as
a prototype of the ("gauge invariant") arborescent vertex \cite{arborgauge}
\be
{\cal V} := \sum_X \
\frac{\left.\bar S_RU_R^{-1}|X\right>\otimes \left<X|U_R\bar S_R\right.
\otimes \left<X|U_R\bar S_R\right.}{ \Lambda'_X}
\ee
or, perhaps,
\be
{\cal V}:= \sum_X  \
\frac{\left.\bar S_R \bar T^{-2}S_RU_R^{-1}|X\right>\otimes
\left<X|U_R\bar S_R\bar T^2 \bar S_R\right.
\otimes \left<X|U_R\bar S_R\bar T^2 \bar S_R\right.}{ \Lambda'_X}
\ee

Matrices $\bar T$ and ${\cal B}$,
as well as the first column $U_{\emptyset Y}=1$ of $U_R$,
are {\it universal} -- do not actually depend on $R$.
What happens is that for a particular $R$
only  finite blocks of these semi-infinite matrices contribute
to the formulas.

Note the presence of quantum dimension $d_R$ in eq.(\ref{arborHOMFLY}),
despite it describes {\it normalized} HOMFLY polynomial,
the un-normalized one would be proportional to $d_R^2$.
As usual in the theory of HOMFLY polynomials parameter $N$ of $SL_N$ gauge algebra
is analytically continued to arbitrary non-integer values and often substituted by $A=q^N$.
Also, we include the combinatorial numbers into $Z$-factors.
In original \cite{M16} they were considered independently, but then \cite{KM17tw}
expressed them explicitly for rectangular representations (thus no need to distinguish and
study them separately in this case),
while for non-rectangular representations it turns impractical to separate them from $Z$ at all.

\bigskip

{\bf Example} $R=[1]$: From
\be
\bar S_{[1]} = \frac{1}{[N]}\left(\begin{array}{cc} 1 & \sqrt{[N+1][N-1]} \\ \sqrt{[N+1][N-1]} & -1
\end{array}\right), \ \ \
\bar T^2 = \left(\begin{array}{cc} 1 & 0 \\ 0 & A^2
\end{array}\right)
\ee
we get
\be
U_{[1]} = \left(\begin{array}{cc} 1 & \sqrt{[N+1][N-1]} \\ 1 &
\frac{A^2-(q^2-1+q^{-2})}{\{q\}^2\sqrt{[N+1][N-1]}}
\end{array}\right)
= \left(\begin{array}{cc} 1 & \sqrt{[N+1][N-1]} \\ 1 &
-\frac{1}{\sqrt{[N+1][N-1]}}+\frac{A}{\{q\}}\frac{[N]}{\sqrt{[N+1][N-1]}}
\end{array}\right), \ \ \
{\cal B} = \left(\begin{array}{ccc} 1 & 0  \\ -A^2 & A^2
\end{array}\right)
\ee
and
\be
Z_{[1]}^\emptyset = 1, \ \ \ \ \ \ Z^{[1]}_{[1]} = \{Aq\}\{A/q\} = D_1D_{-1}= \{q\}^2[N+1][N-1]
\ee
with $\{x\}:=x-x^{-1}$, $D_n:=\{Aq^n\}$,
so that
\be
H_{[1]}^{{\rm twist}_m} = 1 + Z^{[1]}_{[1]}\cdot
\left({\cal B}^{m+1}_{[1],\emptyset}+{\cal B}^{m+1}_{[1],[1]}\right)
= 1 - \{Aq\}\{A/q\}\cdot\sum_{i=1}^{m} A^{2i}
\ee
To compare, before the $U$-"rotation", which converted it into {\it triangular} ${\cal B}$,
the {\it symmetric} matrix was
\be
\bar S_{[1]} \bar T^2 \bar S_{[1]} = \frac{1}{[2] [N]}\left(\begin{array}{cc}
A^2\cdot (q^{-2}[N+1]+q^2[N-1])  & \ \ -A\{q^2\}  \sqrt{[N+1][N-1]}  \\ \\
-A\{q^2\}  \sqrt{[N+1][N-1]}  &  q^{2}[N+1]+q^{-2}[N-1]
\end{array}\right)
\ee
Note that {\bf the first line is $U$ is the same as in $\bar S$}
(consists of the square roots of quantum dimensions of the relevant representations
from $R\otimes \bar R$).
However, while $\bar S$ is finite, some entries of
$U$ are singular in the double-scaling limit when $q,A \longrightarrow 1$
and $N$ is fixed.
This is because $T$ and thus $\bar S\bar T\bar S$ in this limit tend to a unit matrix,
which is preserved by any $U$-rotation, but does not satisfy (\ref{sumB}) --
thus, when approaching the limit, $U$ develops a singularity. $\bullet$

\bigskip

{\bf Example} $R=[2]$: Likewise from
{\footnotesize
\be
\bar S_{[2]} =  \frac{[2]}{[N][N+1]}\left(\begin{array}{ccc}
1 & \sqrt{[N+1][N-1]} & \frac{[N]}{[2]}\sqrt{[N+3][N-1]}\\ \\
\sqrt{[N+1][N-1]} & \frac{[N+1]}{[2][N+2]}\Big([N+3][N-1]-1\Big)
&  -\frac{[N]}{[N+2]}\sqrt{[N+3][N+1]} \\ \\
\phantom.\frac{[N]}{[2]}\sqrt{[N+3][N-1]} & -\frac{[N]}{[N+2]}\sqrt{[N+3][N+1]}
&  \frac{[N]}{[N+2]}
\end{array}\right), \ \ \
\bar T^2 = \left(\begin{array}{ccc} 1 & 0 & 0  \\ 0 & A^2 & 0 \\ 0 & 0 & q^4A^4
\end{array}\right)
\nn
\ee
}

\noindent
it follows that

\bigskip

\centerline{
{\footnotesize
$
U_{[2]} =  \left(\begin{array}{ccc}
1 & \sqrt{[N+1][N-1]} & \frac{[N]}{[2]}\sqrt{[N+3][N-1]} \\ \\
1 & \frac{(q^6+q^4)A^4-(q^8+q^6-q^4+q^2+1)A^2+q^2}{q^4A^2\{q\}\{q^2\}[N+2]}
\sqrt{\frac{[N+1]}{[N-1]}} \!\!\!\!
&   \frac{A^2q^4-q^6+q^4-1}{q^3\{q\}\{q^2\}}\frac{[N]}{[N+2]} \sqrt{\frac{[N+3]}{[N-1]}} \\ \\
1 & \frac{A^2q^4-q^6+q^2-1}{q^3\{q\}^2[N+2]}\sqrt{\frac{[N+1]}{[N-1]}}
& \!\!\!\!\!\!\!\!\!\!\!
\frac{A^4q^{10}-A^2(q^{12}-q^8+q^6+q^4)+(q^{12}-q^{10}-q^8+2q^6-q^2+1)}{q^6\{q\}^3\{q^2\}
[N+2]\sqrt{[N+3][N-1]}}
\end{array}\right), \ \ \
{\cal B} = \left(\begin{array}{ccc} 1 & 0 & 0  \\ -A^2 & A^2 & 0 \\
q^2A^4 & -(q^2+q^4)A^4 & q^4A^4
\end{array}\right)
\nn
$
}}

\bigskip

\noindent
and
\be
Z_{[2]}^\emptyset = 1, \ \ \ \ \ \ \ \ \
Z^{[1]}_{[2]} = [2]\{Aq^2\}\{A/q\} = \{q\}\{q^2\}[N+2][N-1],
\nn \\
Z^{[2]}_{[2]} = \{Aq^3\}\{Aq^2\}\{A\}\{A/q\} = \{q\}^4[N+3][N+2][N][N-1]
\ee
We see that, unlike $\bar S_R$ and $U_R$,
the matrices $\bar T$ and ${\cal B}$ for $R=[2]$ contain those for $R=[1]$
as sub-matrices -- this is a manifestation of their {\it universality}. $\bullet$

\bigskip

These examples have a far-going generalization -- to arbitrary
{\bf rectangular representations $R=[r^s]$}.
Relevant in this case are composite representations
of peculiar diagonal form, which we denote by $X=(\lambda,\lambda)$ or $Y=(\mu,\mu)$ --
only they appear in the product
\be
R=[r^s] \ \ \ \Longrightarrow \ \ \ \
R\otimes \bar R = \oplus_{\lambda\in R}\ (\lambda|\lambda)
\label{rectrepdeco}
\ee
For such diagonal $X,Y$ a general expression is known for ${\cal B}_{XY}$
through the skew Schur functions $\chi_{\lambda/\mu}$, evaluated at the peculiar "unit locus"
$p^\circ_k = \frac{\{q\}^k}{\{q^k\}}\ \stackrel{q\rightarrow 1}{\longrightarrow} \delta_{k,1}$:
\be
{\cal B}_{XY}
= {\cal B}_{(\lambda,\lambda),(\mu,\mu)}
= (-)^{|\lambda|-|\mu|}\cdot
\Lambda_{(\lambda|\lambda)}\cdot \frac{
\chi_{\lambda^\vee/\mu^\vee}^\circ\cdot \chi_\mu^\circ  }{\chi_\lambda^\circ}\,,
\ \  \ \
\boxed{
({\cal B}^{-1})_{XY}
=({\cal B}^{-1})_{(\lambda,\lambda),(\mu,\mu)}
= \frac{ \chi_{\lambda/\mu}^\circ\cdot \chi_\mu^\circ  }{\chi_\lambda^\circ}
\cdot \frac{1}{\Lambda_{(\mu,\mu)}}
}
\label{Bthroughchar}
\ee
where $\lambda^\vee$ denotes the transposition of the Young diagram $\lambda$,
we remind that $\chi_{\lambda^\vee}\{p_k\} = (-)^{|\lambda|}\chi_\lambda\{-p_k\}$.
Inverse matrix ${\cal B}^{-1}$ does not contain transpositions and looks even simpler
than ${\cal B}$ itself.
The evolution eigenvalues $\bar T^2_\lambda = \Lambda_{(\lambda|\lambda)}$
and the trefoil $F$-function
$ \Lambda'_{(\lambda|\lambda)}=F_{(\lambda,\lambda)}^{(1)}$ are monomials,
expressed through the hook parameters of
$\lambda=(a_1,b_1|a_2,b_2|,\ldots)$:
\be
\Lambda_{(\lambda|\lambda)} =
\prod_{i=1}^{\#_{\text{hooks}}(\lambda)}
(q^{2(a_i-b_i)}A^2)^{\overbrace{a_i+b_i+1}^{\text{hook length}}}
= A^{2|\lambda|}\prod_{i=1}^{\#_{\text{hooks}}(\lambda)}
q^{2(a_i-b_i)(a_i+b_i+1)}
\nn \\
\Lambda'_{(\lambda|\lambda)} =
\prod_{i=1}^{\#_{\text{hooks}}(\lambda)}(-q^{a_i-b_i}A^2)^{a_i+b_i+1}
= (-A^2)^{|\lambda|}\prod_{i=1}^{\#_{\text{hooks}}(\lambda)}
q^{(a_i-b_i)(a_i+b_i+1)}
\label{Lambdas}
\ee
(the difference between the two formulas is the minus sign and the lack of $2$
in the degree of $q$  in the second case).
In what follows we abbreviate the notation to
$\Lambda_\lambda:=\Lambda_{(\lambda|\lambda)}=\bar T^2_\lambda$
and $\Lambda'_\lambda:=\Lambda'_{(\lambda|\lambda)}$.
This abbreviation, however, reflects a basic problem -- what is the right
interpretation of states $X=(\lambda|\lambda)$: should they be considered
as $X\in R\otimes \bar R$ or as $\lambda\in R$.
This ambiguity is still unresolved and it will be present in our notation
throughout this text.

\bigskip

If we manage to guess a general formula for $U$,
then $\bar S$ will be directly extractable from (\ref{BviaU}).
Also we can hope to guess what are ${\cal B}$ and $U$ for no-rectangular $R$,
when multiplicities occur and $\bar S$ is not well defined without
additional "gauge-fixing" requirements.

\section{The ${\cal E}$-based approach
\label{Eform}}

Another possibility is just the opposite: if we know ${\cal B}$, then
\be
\boxed{{\cal E}^{-1}:=\bar S_RU_R^{-1}}
\ee
is its diagonalizing matrix, obtained by solving a  linear system
\be
\bar T^2(\bar S_RU_R^{-1}) = (\bar S_RU_R^{-1}){\cal B} \ \  \Longleftrightarrow \ \
{\cal E} \bar T^2    =  {\cal B}{\cal E}
\label{Eeigen}
\ee
Since ${\cal B}$ is triangular and universal, one can expect the same from ${\cal E}$
(though the truth will be a little more involved, see s.\ref{univers} below).
Moreover, the double-braid factorization, discovered in \cite{M16},
allows one to express $\bar S$
through ${\cal E}$ only (and generally-known $\bar T$, $\Lambda'$ and $Z$) --
see eq.(\ref{mystid}) above:
\be
\bar S = \frac{1}{d_R}\cdot \bar T^2 {\cal E}^{\rm tr} \frac{Z}{\Lambda'} {\cal E} \bar T^2
\label{decobSquadr}
\ee
Together with
\be
{\cal B} = {\cal E} \bar T^2 {\cal E}^{-1},
\ \ \ \ \ \
U={\cal E}\bar S
\ee
this provides complete description of the pentad in terms of ${\cal E}$.

Actually, (\ref{decobSquadr}) can serve as expression for $\bar S$ through
${\cal E}=\frac{U\bar S}{d_R}$,  alternative to
\be
\bar S^{[r^s]}_{\mu\nu} = \frac{\chi^*_{[r^s]}}{\sqrt{{\cal D}_\mu{\cal D}_\nu}}
\sum_{\mu,\nu\subset\lambda\subset [r^s]}
Z_\lambda^{[r^s]} f_{\lambda\mu}f_{\lambda\nu}\, ,
\ \ \ \ \ \ \ \ \ \ \ \
f_{\lambda\mu} = {\cal E}_{\lambda\mu}\Lambda_\mu\sum_{\lambda'}{\cal E}^{-1}_{\mu\lambda'}
\label{fvsef}
\ee
Note also that decomposition
\be
\bar T^{-2}\bar S\bar T^{-2} = {\cal E}^{tr} \frac{Z}{\Lambda'} {\cal E}
\ee
is a kind of a dual to expression through the second exclusive Racah matrix $S$
\be
\bar T \bar S \bar T = ST^{-1}S^{-1} \ \ \ \Longleftrightarrow \ \ \ \
\sum_\mu  \bar T_\lambda\bar S_{\lambda\mu }\bar T_\mu S_{\mu\nu} T_\nu
= S_{\lambda\nu}
\label{SfrobS}
\ee
$S$ is the standard diagonalizing matrix for $\bar S$, while ${\cal E}$
diagonalizes it as a quadratic form (i.e. the two decompositions correspond
to treating symmetric $\bar S$ as a tensor with
respectively one covariant and one contravariant index or with two covariant indices.

\section{Rectangular case, examples
\label{exarect}}

For $R=[r]$ the matrix elements of $U$ are:
\be
U_{ \emptyset,[j]} =
\left(\prod_{k=1}^{j-1} \frac{\{Aq^{k-2}\}}{\{q^k\}}\right)\sqrt{\frac{\{Aq^{2j-3}\}}{\{A/q\}}}
\ee
\be
U_{[1],[j]} = U_{ \emptyset,[j]} + \frac{\{q^{j}\}}{A\{Aq^r\}\cdot \{q^r\}}
\left(\prod_{k=1}^{j} \frac{\{Aq^{k-1}\}}{\{q^k\}}\right) \sqrt{\frac{\{Aq^{2j-1}\}}{\{A/q\}}}
\ee
\be
U_{[i],[1]} = U_{\emptyset,[1]} + \frac{\{q^{i}\}}{q^{i-1}\{q\}}\cdot
\frac{\{A\}}{A\{Aq^r\}\cdot\{q^r\}}\cdot\sqrt{\frac{\{Aq\}}{\{A/q\}}}
\ee

In fact, not only ${\cal B} = U\bar S \bar T^2 \bar S U^{-1}= (U\bar S) \bar T^2(U\bar S)^{-1}$,
but its constituent $ U\bar S$ is triangular and universal --
and can be explicitly evaluated.
However, in variance from ${\cal B}$, the entries of ${\cal E} = U\bar S$ depend on $A$
in a rather complicated way.
For symmetric $R=[r]$:
\be
(U\bar S)_{ij} = d_{[r]}\cdot
\frac{(-)^{j-1}q^{\frac{(i-1)(i-2)}{2}-2(j-1)(j-2)}A^{i+1-2j}\{q\}^{ j-1}}
{\prod_{k=0}^{i+j-3} \{Aq^k\}}
\cdot \frac{[i-1]!}{ [i-j]!}
\sqrt{\frac{\{Aq^{2j-3}\}}{\{A/q\}}}
\ee
Because of factorial $[i-j]!$ in denominator this expression vanishes for $j>i$.
It is convenient to introduce a condensed notation
$D_n \, :=\{Aq^n\}$ and $D_n! \, := \prod_{i=0}^n D_i$.
With this definition $D_{-1}!=1$ and in the final formulas we will always write the ratio
$D_n!/D_{-1}!$, which is independent of the lower boundary in the product.

For $R=[1]$, $R=[2]$ and $R=[1,1]$ we have respectively
\be
{\cal E}_{[1]}= U\bar S = d_{[1]}\cdot\left(\begin{array}{cc }
1&0 \\ \frac{A}{D_0}& -\frac{\{q\}}{AD_0\sqrt{D_1D_{-1}}}
\end{array}\right)
\ee
\be
{\cal E}_{[2]}=U\bar S = d_{[2]}\cdot\left(\begin{array}{ccc}
1&0&0 \\ \frac{A}{D_0}& -\frac{\{q\}}{AD_0\sqrt{D_1D_{-1}}} & 0 \\
\frac{qA^2}{D_0D_1} & -\frac{q\{q^2\}}{D_0D_2\sqrt{D_1D_{-1}}} &
\frac{ \{q\}\{q^2\}}{q^3A^2D_0D_1D_2\sqrt{D_3D_{-1}}}
\end{array}\right)
\ee
and
\be
{\cal E}_{[1,1]}=U\bar S = d_{[1,1]}\cdot\left(\begin{array}{ccc}
1&0&0 \\ \frac{A}{D_0}& -\frac{\{q\}}{AD_0\sqrt{D_1D_{-1}}} & 0 \\
\frac{ A^2}{qD_0D_{-1}} & -\frac{ \{q^2\}}{q D_0D_{-2}\sqrt{D_1D_{-1}}} &
\frac{q^3 \{q\}\{q^2\}}{ A^2D_0D_{-1}D_{-2}\sqrt{D_1D_{-3}}}
\end{array}\right)
\ee

For $R=[2,2]$
\be
\frac{{\cal E}_{[2,2]}}{d_{[2,2]}} = \frac{U\bar S}{d_{[2,2]}} =
\label{E22}
\ee

\centerline{{\tiny
$
\cdot\left(\begin{array}{cccccc}
1&0&0&0&0&0 \\ \frac{A}{D_0}& -\frac{\{q\}}{AD_0\sqrt{D_1D_{-1}}} & 0&0&0&0\\
\frac{A^2}{qD_0D_{-1}} & -\frac{\{q^2\}}{qD_0D_{-2}\sqrt{D_1D_{-1}}}&
\frac{q^3\{q\}\{q^2\}}{A^2D_0D_{-1}D_{-2} \sqrt{D_1D_{-3}}}&0&0&0 \\
\frac{qA^2}{D_0D_1 }& -\frac{q\{q^2\}}{D_0D_2\sqrt{D_1D_{-1}}} & 0 &
\frac{\{q\}\{q^2\}}{q^3A^2D_0D_1D_2\sqrt{D_3D_{-1}}} & 0 & 0 \\
\frac{A^3}{ D_1D_0D_{-1}} & -\frac{A\{q^3\}}{D_2D_0D_{-2}\sqrt{D_1D_{-1}}} &
\frac{q^4\{q\}\{q^3\}}{AD_2D_0D_{-1}D_{-2}\sqrt{D_1D_{-3}}} &
\frac{\{q\}\{q^3\}}{q^4AD_2D_1D_0D_{-2}\sqrt{D_3D_{-1}}} &
-\frac{\{q\}^2\{q^3\}}{A^3D_2D_1D_0D_{-1}D_{-2}\sqrt{D_3D_{-3}}}& 0 \\
\frac{A^4}{D_1D_0^2D_{-1}}&
-\frac{A^2[2]\{q^2\}}{D_2D_0^2D_{-2}\sqrt{D_1D_{-1}}}&
\frac{q^4\{q^2\}\{q^3\}}{D_2D_1D_0D_{-1}D_{-2}\sqrt{D_1D_{-3}}} &
\frac{\{q^2\}\{q^3\}}{q^4D_2D_1D_0D_{-1}D_{-2}\sqrt{D_3D_{-1}}}&
-\frac{\{q^2\}^2\{q^3\}}{A^2D_2D_1D_0^2D_{-1}D_{-2}\sqrt{D_3D_{-3}}} &
\frac{\{q\}\{q^2\}^2\{q^3\}}{A^4D_2D_1D_0^2D_{-1}D_{-2}
\sqrt{D_3D_1D_{-1}D_{-3}}}
\end{array}\right)
$
}}

\bigskip

\bigskip

Dimensions of the composite representations,
defining the first line of Racah matrix,
$\bar S_{\emptyset\lambda} = \frac{\sqrt{d_{(\lambda,\lambda)}}}{d_R}$
and partly recognizable also in the entries of ${\cal E}$, are

{\footnotesize
\be
 d_{(\emptyset,\emptyset)}=1, \ \ \ \ \ \ \
 d_{([1],[1])}=D_1D_{-1}, \ \ \ \ \ \ \
 d_{([2],[2])}=\frac{D_{3}D_{0}^2D_{-1}}{[2]^2}, \ \ \ \ \ \ \
 d_{([1,1],[1,1])}=\frac{D_{1}D_{0}^2D_{-3}}{[2]^2}, \ \ \ \ \ \ \
 d_{([2,1],[2,1])}=\frac{D_{3}D_{1}^2D_{-1}^2D_{-3}}{[3]^2},
 \nn\\
 d_{([2,2],[2,2])}=\frac{D_3D_{2}^2D_1D_{-1}D_{-2}^2D_{-3}}{[3]^2[2]^4 }
\nn
\ee}

\bigskip

\section{A universal version of ${\cal E}$
\label{univers}}

Examples in the previous section demonstrate that, against expectation,
${\cal E}$ is not fully {\it universal}:  
it contains factors $d_R$, which explicitly depend on $R$.
This is because
solution to the system (\ref{Eeigen}) does not immediately reproduce
${\cal E}$, instead it is ambiguous -- defined up to right multiplication
by a diagonal matrix,
\be
{\cal E} \ \longrightarrow \ {\cal E}\cdot {\cal D}
\label{Etrans}
\ee
In the true ${\cal E}$ the freedom is fixed by the request that $U_{\lambda \emptyset}=1$,
which is important to reproduce the $Z$-factors:
\be
U_{X\emptyset} = 1 \ \ \ \ \Longrightarrow \ \ \ \
Z_R^X = d_R \cdot
\Big(\bar S_R \bar T^{2} \bar S_R \bar T^{-2} \bar S_R U_R^{-1} \Big)_{\emptyset X}
= d_R \cdot\sum_Y
\Big(\bar S_R \bar T^{2} \bar S_R \bar T^{-2}\Big)_{\emptyset Y}   {\cal E}^{-1}_{YX}
\ee
This is not quite a simple condition on the diagonal matrix ${\cal D}$ in (\ref{Etrans}).
One way to express it is through the sum rules for every given sub-representation $\lambda\subset R$:
\be
\sum_{\mu\subset \lambda}
{\cal E}_{\lambda\mu}\Lambda_\mu \bar S_{\emptyset \mu} = \delta_{\lambda,\emptyset},
\ \ \ \ \ \ \ \ \ \
\sum_{\mu\subset \lambda}
{\cal E}_{\lambda\mu}\Lambda_\mu^2 \bar S_{\emptyset \mu}  =   \Lambda'_\lambda,
\label{srules}
\ee
for example, for $R=[1]$,
\be
\left(\begin{array}{cc} 1 & 0 \\ \frac{A}{D_0} & -\frac{\{q\}}{AD_0\sqrt{D_1D_{-1}}}
\end{array}\right)\left(\begin{array}{c} 1 \\ \frac{\sqrt{D_1D_{-1}}}{\{q\}}\Lambda_1
\end{array}\right) =
\left(\begin{array}{c} 1 \\ \frac{A-\frac{\Lambda_1 }{A}}{D_0} \end{array}\right)
=
\left(\begin{array}{c} 1 \\ \frac{A-A}{D_0} \end{array}\right)
= \left(\begin{array}{c} 1 \\ 0 \end{array}\right)
\ee
and
\be
\left(\begin{array}{cc} 1 & 0 \\ \frac{A}{D_0} & -\frac{\{q\}}{AD_0\sqrt{D_1D_{-1}}}
\end{array}\right)\left(\begin{array}{c} 1 \\ \frac{\sqrt{D_1D_{-1}}}{\{q\}}\Lambda_1^2
\end{array}\right) =
\left(\begin{array}{c} 1 \\ \frac{A-\frac{\Lambda_1^2}{A}}{D_0} \end{array}\right)
=
\left(\begin{array}{c} 1 \\ \frac{A-A^3}{D_0} \end{array}\right)
= \left(\begin{array}{c} 1 \\ -A^2 \end{array}\right)
\ee

\bigskip

We can fix the ambiguity (\ref{Etrans}) in an alternative way -- 
by requiring that diagonal elements are unities.
The resulting matrix is not exactly ${\cal E}^{-1}:=\bar S_RU_R^{-1}$,
but it is fully {\it universal}, i.e. independent of representation $R$, like ${\cal B}$:
\be
\check {\cal E}^{-1} =  \left(\begin{array}{c|ccccc}
&\emptyset & [1] & [2] & [3] & \ldots \\
\\
\hline
\\
\emptyset&1 & 0 & 0 &0& \\ \\
\phantom.[1] & -\frac{A}{\{A\}}& 1 & 0& 0 & \ldots\\ \\
\phantom.[2] &\frac{qA^2}{\{Aq^2\}\{Aq\}} & -\frac{q[2]A}{\{Aq^2\}} & 1&0& \\ \\
\phantom.[3] &-\frac{q^3A^3}{\{Aq^4\}\{Aq^3\}\{Aq^2\}} & \frac{q^3[3]A^2}{\{Aq^4\}\{Aq^3\}}
& -\frac{q^2[3]A}{\{Aq^4\}} & 1 \\ \\
\ldots
\end{array}\right)
\label{Esymm}
\ee
In general
the properly normalized ${\cal E}_{\mu\nu}$
(which gives rise to $U_{\mu\emptyset} = 1 \ {\rm or}\ 0$)
differs from $\check{\cal E}$ with
unit non-vanishing diagonal elements, $\check {\cal E}_{\mu\mu}=1 \ {\rm or}\ 0$,
by a rather sophisticated Abelian factor $K$:
\be
{\cal E}_{\mu\nu} = \check{\cal E}_{\mu\nu}\cdot K_\nu
\nn \\
K_\nu = \frac{(-\{q\})^{|\nu|} }{\Lambda_\nu^{1/2}\chi^0_\nu}\cdot
\prod_{hooks\in \nu} \frac{D_{-2b_h-1}!}{D_{2a_h}!}\cdot\frac{1}{\sqrt{D_{2a_h+1}D_{-2b_h-1}}}
\nn\\
K_\nu^{-1}= \Big(-\{q\}\Big)^{-|\nu|} \cdot {\Lambda_\nu^{1/2}\chi^0_\nu}\cdot
 \prod_{hooks\in \nu} \frac{D_{2a_h}!}{D_{-2b_h-1}!}\cdot {\sqrt{D_{2a_h+1}D_{-2b_h-1}}}
\ee

\noindent
In hook variables the Young diagram $\nu=(a_1,b_1|a_2,b_2|\ldots)$,
e.g. $\nu=[5,4,1,1]=(4,3|2,0)$.
Comparing (\ref{Esymm}) with the similar examples of $U$ and ${\cal E}$ in the previous sections,
we see, that $\check{\cal E}^{-1}$ can be simpler to find in a general form.

\section{Non-rectangular case, generalities}

The main difficulty when one passes from rectangular to non-rectangular representations $R$
is that
\be
R\otimes\bar R = \sum_X c_R^X\cdot X
= \sum_{\lambda,\lambda'\in R} c_R^{\lambda\lambda'}(\lambda,\lambda')
\label{genrepdeco}
\ee
now contains a whole variety of composite representations
$X=(\lambda,\lambda')$, where, first, $\lambda'$ can be different from $\lambda$ and, second,
they can come with non-trivial multiplicities $c_{R}^{\lambda\lambda'}>1$.
This makes exclusive Racah matrix $\bar S$ and other members of the pentad
$(\bar S,S,{\cal B}, U,{\cal E})$  much bigger and more complicated.
For example, for $R=[2,1]$
\be
\phantom. [2,1]\otimes \overline{[2,1]} = (\emptyset,\emptyset) + 4\cdot ([1],[1]) +
([2],[2]) + ([1,1],[1,1]) + \underline{([2],[1,1])+([1,1],[2])} + ([2,1],[2,1])
\label{deco21}
\ee
and all matrices are $10\times 10$.
However, as explained in \cite{M19nr}, one can actually handle twist knot polynomials
in terms of just $6\times 6$ "reduced" matrices ${\cal B}^{\rm red}$,
with all the four $([1],[1])$ shrinked to one,
and two underlined non-diagonal representations also substituted by one.
This, however, is not enough for description of Racah matrices and thus for
handling arbitrary arborescent knots by the technique of \cite{arbor}.
The brute-force extraction of Racah matrices from twist-knot calculus is described
in \cite{MnonrectRacah} for representation $R=[3,1]$,
and it is quite difficult to use for bigger non-rectangular $R$.
Clearly, we need a more systematic approach to the problem.

As we explain in this paper, the true situation is more interesting.
For generic non-rectangular representation $R$ symmetric Racah matrix
$\bar S_R$ can have blocks of the type
\be
\left(\begin{array}{cc} 0 & 1 \\ 1 & 0 \end{array}\right)
= \frac{1}{2i} \left(\begin{array}{cc} 1 & i \\ i & 1 \end{array}\right)^2
\label{sigma1asasquare}
\ee
which are separated from the relevant part of the matrix $\bar S^{\rm rel}$
and do not actually contribute
to arborescent knot calculus, because the corresponding $Z$-factors are vanishing.
In above example of $R=[2,1]$ the "relevant" part $\bar S^{\rm rel}$
is $8\times 8$, and the decoupling $2\times 2$ block consists of a
different $([2],[1,1])-([1,1],[2])$ and a single linear combination of the four
copies of $([1],[1])$.
The further reduction ${\cal B}^{\rm rel} \ \longrightarrow \ {\cal B}^{\rm red}$
from $8\times 8$ to $6\times 6$ is made possible by the vanishing of one extra
$Z$-factor for the three remaining $([1],[1])$ and by existence of a simple
combination of the other two -- which causes the $Z^{[1]}$-factor to be non-factorized
in the $6\times 6$ formalism (while it is factorized, as all other $Z$-factors)
when we stay with the $8\times 8$ matrices).
One more general remark is that because of
(\ref{sigma1asasquare}) in {\it irrelevant} $2\times 2$ sectors the
${\cal E}$ is not triangular, but rather has the form
\be
\frac{1}{\sqrt{2}}\left(\begin{array}{cc} \frac{1}{\sqrt{i}} & \sqrt{i} \\ \\
\sqrt{i} & \frac{1}{\sqrt{i}} \end{array}\right)
\ee
In what follows we ignore these irrelevant pieces and identify ${\cal E}$
with triangular ${\cal E}^{\rm rel}$.

In what follows we are going to describe all these structures in some detail
for the case of $R=[2,1]$ and provide the next example of $R=[3,1]$,
but only for the ${\cal B}$ matrix.
We also present the decompositions  like (\ref{deco21})
for some more complicated non-rectangular representations $R$ --
also a necessary step for further generalizations.

\section{Reduced version of ${\cal B}$
\label{redB}}

As already mentioned, one can use the technique
of this paper for two purposes:
calculation of knot polynomials for twist knots and calculation of Racah matrices.
For non-rectangular representations the first purpose
is significantly simpler, because for twist knot polynomials
we need a little less than $\bar S$ --
just its particular matrix elements,
which are captured by the reduced matrix ${\cal B}^{\rm red}$.
This was already explained in \cite{M19nr},
and we briefly repeat this description to make the story complete.
Like in \cite{M19nr} we consider the simplest case of $R=[r,1]$,
for which the answers are already known from \cite{Mnonrect}.

In this case in addition to the $2r+1$ diagrams $X=(\lambda,\lambda)$ with
$\lambda \subset R=[r,1]$, i.e. $\lambda =\emptyset, [i], [i,1]$, $i=1,\ldots,r$
there are $r-1$ additional composite pairs  $\boxed{\tilde X_i=([i-1,1],[i])\oplus([i],[i-1,1])}$
with the same dimensions and eigenvalues
\vspace{-0.3cm}
\be
\tilde\Lambda_i = (q^{i-2}A)^{2i}, \ \ \ {i=2,\ldots,r}
\ee
each contributing {\it once} to the differential expansion.
These $\tilde X_i$ contribute $r-1$ additional lines to the
reduced matrix ${\cal B}^{\rm red}$,
which thus becomes of the size $2r+1+r-1=3r$.
Remarkably, ${\cal B}^{\rm red}$ remains {\it triangular}, though a notion of ordering for
generic composites $X$, appearing in (\ref{genrepdeco}),
gets somewhat more subtle than (\ref{rectrepdeco}).
The first $2r+1$ lines remain as they were in (\ref{Bthroughchar}).
The new entries in the new $r-1$ lines $\tilde X_i$ with $i=2,\ldots,r$ are:

\be
\boxed{
\begin{array}{ccl}
{\cal B}^{\rm red}_{\tilde X_i,\emptyset} = \ \ \ \
&  \frac{(-)^{i+1} \tilde\Lambda_i }{ q^{(i-1)(i-2)}}\cdot A^2
\nn\\ \nn\\
\phantom.\! {\cal B}^{\rm red}_{\tilde X_i,[j]} = \ \ \ \
&\frac{(-)^{i+j-1}\tilde\Lambda_i}{q^{(i-1)(i-j)} }\cdot
\frac{[i-2]!}{[i-j]![j-1]!} \cdot
\frac{ [i-1]\cdot q^{3i+j-2} A^2
- [i-j] \cdot q^{i-3} A^2-   [j-1] }{q^{2j}-1} \ \ \ \ & _{j=1,\ldots, i}
\nn\\ \nn\\
{\cal B}^{\rm red}_{\tilde X_i,[j,1]} = \ \ \ \  &\frac{(-)^{i+j-1}\tilde\Lambda_i}{q^{i^2-ij-2i-j+7} }
\cdot \frac{[i-2]!}{[i-j-1]![j-1]!} \cdot
\frac{(A^2-q^2)(A^2-q^6)}{(q^{2j+2}-1)(A^2q^{2j-4}-1)} & _{j =1,\ldots, i-1}
\nn\\ \nn\\
{\cal B}^{\rm red}_{\tilde X_i,\tilde X_j} = \ \ \ \
& \frac{(-)^{i+j}\tilde\Lambda_i}{q^{(i-1)(i-j)} }\cdot
\frac{[i-2]!}{[i-j]![j-2]!}\cdot\frac{A^2q^{2i-4}-1 }{A^2q^{2j-4}-1}
& _{j =2, \ldots, i} \\
\end{array}
}
\label{newB}
\ee

\bigskip

\noindent
In particular,
\vspace{-0.3cm}
\be
{\cal B}^{\rm red}_{\tilde X_i, [1]} =
\ \ & (-)^i \tilde\Lambda_i \cdot \frac{[i+1]\cdot A^2  }{q^{(i-2)^2}}
\nn \\ 
{\cal B}^{\rm red}_{\tilde X_i,[1,1]} = \ \
&  (-)^i \tilde\Lambda_i\cdot \frac{ A^2 -q^6}{q^{i^2-3i+4}\cdot(q^{4}-1)}
\nn \\ 
{\cal B}^{\rm red}_{\tilde X_i,[i]} = \ \ & -\tilde\Lambda_i\cdot \frac{ A^2q^{4i-2}-1   }{q^{2i}-1}
\nn \\ 
{\cal B}^{\rm red}_{\tilde X_i,[i,1]} = \ \ & 0
\ee

\be
\boxed{
\begin{array}{ccl}
{\cal B}^{\rm red}_{\tilde X_i,\emptyset} = \ \ \ \  &  {\cal B}_{[i,1],\emptyset}
\nn\\ \nn\\
{\cal B}^{\rm red}_{\tilde X_i,[j]} = \ \ \ \
& {\cal B} _{[i,1],[j]} +
\frac{(-)^{i+j-1}}{q^{i^2+ij-2i-2j-2} }\cdot
\frac{[i-2]!}{[i-j]![j-2]!} \cdot
\frac{  A^{2i+1}D_{i-2}}{\{q^j\}} \ \ \ \
& _{j=1,\ldots, i}
\nn\\ \nn\\
{\cal B}^{\rm red}_{\tilde X_i,[1,1]} = \ \ \ \
&  {\cal B}_{[i,1],\emptyset} + (-)^iq^{i^2-2}A^{2i+1}\cdot \frac{D_{i-2}}{\{q^2\}}
\nn\\ \nn\\
{\cal B}^{\rm red}_{\tilde X_i,[j,1]} = \ \ \ \
& (-)^{i+j+1}q^{i^2+ij-2i-j-2}A^{2i+1}\cdot \frac{[i-2]!}{[i-j-1]![j-1]!}
\cdot \frac{D_{-1}D_{-3}}{\{q^{j+1}\}D_{j-2}}
& _{j =1,\ldots, i-1}
\nn\\ \nn\\
{\cal B}^{\rm red}_{\tilde X_i,[i,1]} = \ \ \ \
& 0
\nn\\ \nn\\
{\cal B}^{\rm red}_{\tilde X_i,\tilde X_j} = \ \ \ \
&  (-)^{i+j}A^{2i}q^{(i-2)(i+j)} \cdot
\frac{[i-2]!}{[i-j]![j-2]!}\cdot\frac{D_{i-2}}{D_{j-2}}
& _{j =2, \ldots, i} \\
\end{array}
}
\label{newB1}
\ee

In the simplest case of $R=[2,1]$ the matrix is
\be
{\cal B}^{\rm red}_{[2,1]} = \left(\begin{array}{c|cccccc}
& \emptyset & [1] & [1,1] & [2] & [2,1] & \tilde X_{2} \\
&&&&&&\\
\hline
&&&&&&\\
\phantom.\emptyset & 1 & 0 & 0 & 0 & 0 & 0 \\
&&&&&&\\
\phantom.[1] & -A^2 & A^2 & 0 & 0 & 0 & 0\\
&&&&&&\\
\phantom.[1,1] & \frac{A^4}{q^2} & -\frac{[2]A^4}{q^3} & \frac{A^4}{q^4} & 0 & 0 & 0 \\
&&&&&&\\
\phantom.[2] & q^2A^4 & -[2]q^3A^4 & 0 & q^4A^4 & 0 & 0 \\
&&&&&&\\
\phantom.[2,1] & -A^6 & [3]A^6 & -\frac{[3]A^6}{[2]q} & -\frac{[3]qA^6}{[2]} & A^6 & 0 \\
&&&&&&\\
\tilde X_2 & -A^6 & [3]A^6 & \frac{ (A^2-q^6)A^4}{q^2(q^4-1)}
& -\frac{(A^2q^6-1)A^4}{q^4-1}&0&A^4
\end{array}\right)
\label{Bred21}
\ee

\bigskip

\noindent
The new one -- revealed by consideration of the non-rectangular $R$ -- is the last line.\\

For  $R=[3,1]$ the line $\tilde X_2$ remains the same
-- this is the {\it universality} property of ${\cal B}$ --
and there is one more  line for $\tilde X_3$,  {\it new}  as compared to (\ref{Bthroughchar}):

\bigskip

\centerline{{\footnotesize
$
\begin{array}{c||ccccccccc}
&\emptyset&[1]&[1,1]&[2]&[2,1]&[3]&[3,1]&\tilde X_2&\tilde X_3 \\
&&&&&&&&&\\
\hline
&&&&&&&&&\\
\tilde X_3 & q^4A^8 & -[4]q^5A^8 & -\frac{q^2(A^2-q^6)A^6}{q^4-1} &
\frac{q^4\big(A^2(q^{10}+q^8-1)-1\big)A^6}{q^4-1} & \frac{q^4(A^2-q^2)(A^2-q^6)A^6}{(q^6-1)(A^2-1)}
& -\frac{q^6(A^2q^{10}-1)A^6}{q^6-1} & 0 & -\frac{q^4(A^2q^2-1)A^6}{A^2-1}& q^6A^6
\end{array}
$
}}

\bigskip

\bigskip

As another manifestation of universality,
the matrix ${\cal E}$ for $R=[2,1]$  is the same as (\ref{E22}) for $R=[2,2]$,
except for the very last line, associated with
the composite representation $([2],[1,1])\oplus([1,1],[2])$ instead of
$([2,2],[2,2])$:

\be
\frac{{\cal E}^{\rm red}}{d_{[2,1]}} = \frac{U^{\rm red}\bar S^{\rm red}}{d_{[2,1]}} =
\label{E21}
\ee
\centerline{{\footnotesize
$
\cdot\left(\begin{array}{cccccc}
1&0&0&0&0&0 \\ \frac{A}{D_0}& -\frac{\{q\}}{AD_0\sqrt{D_1D_{-1}}} & 0&0&0&0\\
\frac{A^2}{qD_0D_{-1}} & -\frac{\{q^2\}}{qD_0D_{-2}\sqrt{D_1D_{-1}}}&
\frac{q^3\{q\}\{q^2\}}{A^2D_0D_{-1}D_{-2} \sqrt{D_1D_{-3}}}&0&0&0 \\
\frac{qA^2}{D_0D_1 }& -\frac{q\{q^2\}}{D_0D_2\sqrt{D_1D_{-1}}} & 0 &
\frac{\{q\}\{q^2\}}{q^3A^2D_0D_1D_2\sqrt{D_3D_{-1}}} & 0 & 0 \\
\frac{A^3}{ D_1D_0D_{-1}} & -\frac{A\{q^3\}}{D_2D_0D_{-2}\sqrt{D_1D_{-1}}} &
\frac{q^4\{q\}\{q^3\}}{AD_2D_0D_{-1}D_{-2}\sqrt{D_1D_{-3}}} &
\frac{\{q\}\{q^3\}}{q^4AD_2D_1D_0D_{-2}\sqrt{D_3D_{-1}}} &
-\frac{\{q\}^2\{q^3\}}{A^3D_2D_1D_0D_{-1}D_{-2}\sqrt{D_3D_{-3}}}& 0 \\
\frac{A^3}{D_1D_0D_{-1}}&
-\frac{A[4]\{q\}}{[2]D_2D_0D_{-2}\sqrt{D_1D_{-1}}}&
-\frac{q^4 }{[2]AD_0D_{-1}D_{-2}}\sqrt{\frac{D_{-3}}{D_1}} &
-\frac{1}{q^4[2]AD_2D_1D_0}\sqrt{\frac{D_3}{D_{-1}}}&
0 &
\frac{\sqrt{2}}{[2]AD_0\sqrt{D_2D_1D_{-1}D_{-2}}}
\end{array}\right)
$
}}

\bigskip

The next step after \cite{Mnonrect} and \cite{M19nr} in direction of this section
would be development of twist-knot calculus for arbitrary non-rectangular representations,
i.e. fixing the notion of reduced  and finding its non-trivial matrix elements.
In the next section we just described the decomposition of more complicated representations,
leaving construction of associated ${\cal B}^{red}$ for the future work.

\section{Decomposition of $R\otimes \bar R$}

For arbitrary $N$ we have the following decomposition of the product --
either of representations or of the corresponding characters (Schur functions)
\be
\phantom.[r]\otimes [r^{N-1}] = \oplus_{i=0}^r [N+i,r^{N-2},r-i]
\ \ \ \longrightarrow \ \ \
\chi_{[r]}\{p\} \cdot \chi_{[r^{N-1}]}\{p\} = \sum_{i=0}^r \chi_{[N+i,r^{N-2},r-i]}\{p\}
\ee
Representations appearing at the r.h.s. are called composite

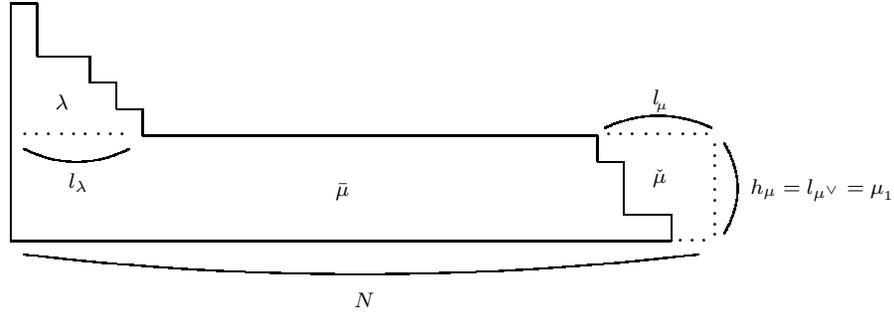
\begin{picture}(300,220)(-90,-110)

\put(0,0){\line(0,1){90}}
\put(0,0){\line(1,0){250}}
\put(50,40){\line(1,0){172}}

\put(0,90){\line(1,0){10}}
\put(10,90){\line(0,-1){20}}
\put(10,70){\line(1,0){20}}
\put(30,70){\line(0,-1){10}}
\put(30,60){\line(1,0){10}}
\put(40,60){\line(0,-1){10}}
\put(40,50){\line(1,0){10}}
\put(50,50){\line(0,-1){10}}

\put(265,2){\mbox{$\vdots$}}
\put(265,15){\mbox{$\vdots$}}
\put(265,28){\mbox{$\vdots$}}

\put(252,0){\mbox{$\ldots$}}
\put(253,40){\mbox{$\ldots$}}
\put(239,40){\mbox{$\ldots$}}
\put(225,40){\mbox{$\ldots$}}

\put(222,40){\line(0,-1){10}}
\put(222,30){\line(1,0){10}}
\put(232,30){\line(0,-1){20}}
\put(232,10){\line(1,0){18}}
\put(250,0){\line(0,1){10}}

\put(0,90){\line(1,0){10}}
\put(10,90){\line(0,-1){20}}
\put(10,70){\line(1,0){20}}
\put(30,70){\line(0,-1){10}}
\put(30,60){\line(1,0){10}}
\put(40,60){\line(0,-1){10}}
\put(40,50){\line(1,0){10}}
\put(50,50){\line(0,-1){10}}

{\footnotesize
\put(123,17){\mbox{$ \bar \mu$}}
\put(17,50){\mbox{$\lambda$}}
\put(243,22){\mbox{$\check \mu$}}
\qbezier(270,3)(280,20)(270,37)
\put(280,18){\mbox{$h_\mu = l_{\mu^{\vee}}=\mu_{_1}$}}
\qbezier(5,-5)(132,-20)(260,-5)
\put(130,-25){\mbox{$N $}}
\qbezier(5,35)(25,25)(45,35)
\put(22,20){\mbox{$l_\lambda$}}
\qbezier(225,43)(245,52)(265,43)
\put(243,52){\mbox{$l_{\!_\mu}$}}
}

\put(4,40){\mbox{$\ldots$}}
\put(18,40){\mbox{$\ldots$}}
\put(32,40){\mbox{$\ldots$}}

\put(-20,-50){\mbox{
$(\lambda,\mu)_N= \Big[\lambda_1+\mu_1,\ldots,\lambda_{l_\lambda}+\mu_1,\
\underbrace{\mu_1,\ldots,\mu_1}_{N-l_{\!_\lambda}-l_{\!_\mu}},\
\mu_1-\mu_{_{l_{\!_\mu}}},\ldots,\mu_1-\mu_2\Big]$
}}

\put(-0,-90){\mbox{
$\chi_{(\lambda,\mu)}[{\cal X}]
=\det{\cal X}^{|\mu|}\cdot\sum_\eta (-)^{|\eta|} \cdot \chi_{\lambda/\eta}[{\cal X}]\cdot
\chi_{\mu/\eta^\vee}[{\cal X}^{-1}]$
}}

\end{picture}

\noindent
and we can rewrite the statement as
\be
\chi_{[r]}\{p\}\cdot\chi_{(0,[r])}\{p\} = \sum_{i=0}^r \chi_{([i],[i])}\{p\}
\label{chirrp}
\ee
where the $i=0$ terms at the r.h.s. is understood as $\chi_{[r^n]}\{p\}$,
not $\chi_{\emptyset}\{p\}=1$.
Here and almost everywhere below we are suppressing the index $N$ --
but remember that composite representations
are explicitly $N$-dependent.
The next step is to restrict $p$-variables to the Miwa locus
$p_k=\sum_{a=1}^N x_a^k=\tr {\cal X}^k$
with the matrix ${\cal X}$ of the same size $N$.
The corresponding restriction of Schur functions is denoted by square brackets:
$\chi_R\{p_k=\tr {\cal X}^k\} = \chi_R[{\cal X}]$.
Immediate corollaries are the simple expression for conjugate representations
\be
\chi_{_{(\mu,\lambda)}}[{\cal X}] = \det{\cal X}^{ \lambda_1 +\mu_1}\cdot
\chi_{_{(\lambda,\mu)}}[{\cal X}^{-1}]
\ee
($\lambda_1$ is the length of the longest row in $\lambda$)
and vanishing of $\chi_\lambda[{\cal X}]$ for Young diagrams $\lambda$ with more than
$N$ lines, $l_\lambda>N$.
Also, one can eliminate any full line of length $N$, but multiply by an extra factor of
$\det{\cal X}$.
For the particular case of symmetric representations we now get:
\be
\det{\cal X}^r\cdot \chi_{_{[r]}}[{\cal X}]\cdot \chi_{_{[r]}}[{\cal X}^{-1}]
= \sum_{i=0}^r \chi_{_{([i],[i])}}[{\cal X}]
\label{chirrX}
\ee

Expression like (\ref{chirrp}) through composite representations are not
obligatory existing.
The simplest example appears already for $R=[2,1]$:
{\footnotesize
\be
\!\!\!\!\!  \!\!\!\!\!  \!\!\!\!\!  \!\!\!\!\!
\begin{array}{rl}
\phantom.[2,1]^2 = &
\phantom. [2,2,2]+\underline{2\cdot[3,2,1]+[4,2]+\underline{[4,1,1]+[3,3]}}
+{[3,1,1,1]+[2,2,1,1]}
\nn \\ \nn \\
\phantom.[2,1]\otimes [2,2,1] = &
\phantom.[2,2,2,2]+\underline{ 2\cdot [3,2,2,1] + [4,2,2]+[3,3,1,1]+[4,3,1]
+ \underline{[4,2,1,1]+[3,3,2]}}
+  {[3,2,1,1,1]+[2,2,2,1,1]}
\nn \\ \nn \\
\phantom.[2,1]\otimes \underbrace{\overline{[2,1]}}_{(\emptyset,[2,1])} = &
[2^N]+\underline{2\cdot ([1],[1])_N+([2],[2])_N+([1,1],[1,1])_N+([2,1],[2,1])_N
+ \underline{([2],[1,1])_N+([1,1],[2])_N}}
+    {[3,2^{N-3},1^3]+[2^{N-1},1,1]}
\end{array}
\nn
\ee
}

\noindent
The first two lines are particular examples of the third one at $N=3$ and $N=4$.
Underlined items can be treated as composite representations,
the very first item is a rather innocent deviation, but the last two
are more serious -- they contain a line of the length $N+1$.
Both problems can be eliminated by rewriting the representation product in terms
of Schur functions and restricting to the Miwa locus:
\be
\det{\cal X}^2\cdot \chi_{[2,1]}[{\cal X}]\cdot \chi_{[2,1]}[{\cal X}^{-1}]
= \det{\cal X}^2  + 2\chi_{([1],[1])}[{\cal X}] + \chi_{([2],[2])}[{\cal X}]
+ \chi_{([1,1],[1,1])}[{\cal X}] + \chi_{([2,1],[2,1])}[{\cal X}]
+ \nn \\ + \chi_{([2],[1,1])}[{\cal X}]
+ \chi_{([1,1],[2])}[{\cal X}] = \ \ \ \
\sum_{\lambda,\lambda'\subset [2,1]} c_{\lambda,\lambda'}^{[2,1]}\cdot
\delta_{|\lambda|,|\lambda'|}\cdot\chi_{(\lambda,\lambda')}[{\cal X}] \ \ \ \
\ee
The two non-underlined terms disappeared, moreover the two non-diagonal
composites in the second line, which were double-underlined,
are now equal to each other.
The formula is further simplified if Miwa locus is further restricted
to $\det{\cal X}=1$ (i.e. from $GL_N$ to $SL_N$).
The resulting expression is very similar to (\ref{chirrX}),
only in general the sum is diagonal only in the {\it size} of the sub-diagrams
and there can be non-trivial multiplicities $c^R_{\lambda,\lambda'}$:
\be
\chi_R[{\cal X}]\cdot\chi_R[{\cal X}^{-1}] =
\sum_{\stackrel{\lambda,\lambda'\subset R}{|\lambda|=|\lambda'|}}
c_{\lambda,\lambda'}^{R}\cdot
\chi_{(\lambda,\lambda')}[{\cal X}] \ \ \ \ \ \ \ \ \ \ \ {\rm for} \ \ \det{\cal X}=1
\ee
This formula is a kind of a dual or inverse to the Kojke formula for the composite
Schur functions \cite{Koj,MMhopf}
\be
\chi_{(\lambda,\mu)}[{\cal X}]
=\det{\cal X}^{|\mu|}\cdot\sum_\eta (-)^{|\eta|} \cdot \chi_{\lambda/\eta}[{\cal X}]\cdot
\chi_{\mu/\eta^\vee}[{\cal X}^{-1}]
\ee
where multiplicities do not show up explicitly:
they are hidden/incorporated in  the skew characters.

We now describe these multiplicities in some particular cases.
To make formulas better readable we omit ${\cal X}$ and write
$\overline{\chi_R}=\chi_R[{\cal X}^{-1}]$.

\begin{itemize}

\item{For arbitrary rectangular $R=[r^s]$ there are only diagonal contributions,
$\lambda'=\lambda$, all with multiplicities one:
\be
\chi_{[r^s]} \cdot\overline{\chi_{[r^s]}}  =
\sum_{\lambda\subset [r^s]}
\chi_{(\lambda,\lambda)}
\label{chirect}
\ee
}

\item{For the simplest double-line diagrams $R=[r,1]$ non=diagonal terms are present,
and single-line sub-diagrams, besides two, come with multiplicities $2$
\be
\chi_{[r,1]} \cdot\overline{\chi_{[r,1]}} =
1+2\sum_{i=1}^{r-1}\chi_{([i],[i])}+\chi_{([r],[r])}
+\sum_{i=1}^{r} \chi_{([i,1],[i,1])}
+\sum_{i=1}^{r-1}\Big(\chi_{([i+1],[i,1])}+\chi_{([i,1],[i+1])}\Big)
\ee
The two items in the last sum are pairwise equal.
In the particular case of $r=1$ we get just the particular case of (\ref{chirect}):
$\chi_{[1,1]} \cdot\overline{\chi_{[1,1]}} =
1+\chi_{([1],[1])}+\chi_{([1,1],[1,1])}$.
}

\item{For the next -- two-line  -- example $R=[r,2]$ the decomposition gets a little trickier
(diagonal composites are collected in  the first two lines):
\be
\chi_{[r,2]} \cdot\overline{\chi_{[r,2]}} =
1+ \left(-\chi_{_{([1],[1])}}+3\sum_{i=1}^{r-2} \chi_{_{([i],[i])}}
 +2\chi_{_{([r-1],[r-1])}}+\chi_{_{([r],[r])}}\right)+\nn\\
+\left(\chi_{_{([1,1],[1,1])}}+2\sum_{i=2}^{r-1}  \chi_{_{([i,1],[i,1])}}
+ \chi_{_{([r,1],[r,1])}}\right)
+\sum_{i=2}^{r} \chi_{_{([i,2],[i,2])}}
+ \nn \\ \!\!\!\!\!\!\!\!\!\!\!\!\!\!\!\!
+\left(\Big(\chi_{_{([2],[1,1])}}+\chi_{_{([1,1],[2])}}\Big)
+2\sum_{i=2}^{r-2} \Big(\chi_{_{([i+1],[i,1])}}+\chi_{_{([i,1],[i+1])}}\Big)
+\Big(\chi_{_{([r],[r-1,1])}}+\chi_{_{([r-1,1],[r])}}\Big)\right)\cdot(1-\delta_{r,2})
+\nn\\
+\sum_{i=2}^{r-1}\Big(\chi_{_{([i+1,1],[i,2])}}+\chi_{_{([i,2],[i+1,1])}}\Big)
+\sum_{i=2}^{r-2}\Big(\chi_{_{([i+2],[i,2])}}+\chi_{_{([i,2],[i+2])}}\Big) \ \ \ \ \ \ \
\ee
in particular,
\be
\chi_{_{[2,2]}}\cdot\overline{\chi_{_{[2,2]}}} = 1+ \chi_{_{([1],[1])}}
+ \chi_{_{([2],[2])}} + \chi_{_{([1,1],[1,1])}}
+ \chi_{_{([2,1],[2,1])}} + \chi_{_{([2,2],[2,2])}}
\nn \\ \nn \\
\chi_{_{[3,2]}}\cdot\overline{\chi_{_{[3,2]}}} = 1+ 2\cdot\chi_{_{([1],[1])}}
+ 2\cdot \chi_{_{([2],[2])}} + \chi_{_{([1,1],[1,1])}}
+ \chi_{_{([2],[1,1])}}+ \chi_{_{([1,1],[2])}}
 + \nn\\
 + \chi_{_{([3],[3])}}+2\cdot\chi_{_{([2,1],[2,1])}}
 + \chi_{_{([3],[2,1])}}+ \chi_{_{([2,1],[3])}}
  + \nn \\
+ \chi_{_{([3,1],[3,1])}}+ \chi_{_{([2,2],[2,2])}}
+ \chi_{_{([3,1],[2,2])}}+ \chi_{_{([2,2],[3,1])}}
 + \chi_{_{([3,2],[3,2])}}
\nn \\ \nn \\
\chi_{_{[4,2]}}\cdot\overline{\chi_{_{[4,2]}}} = 1+ 2\cdot\chi_{_{([1],[1])}}
+ 3\cdot \chi_{_{([2],[2])}} + \chi_{_{([1,1],[1,1])}}
+ \chi_{_{([2],[1,1])}}+ \chi_{_{([1,1],[2])}}
 + \nn\\
 +2\cdot \chi_{_{([3],[3])}}+2\cdot\chi_{_{([2,1],[2,1])}}
 + 2\cdot\chi_{_{([3],[2,1])}}+ 2\cdot\chi_{_{([2,1],[3])}}
  + \nn \\
 + \chi_{_{([4],[4])}} + 2\cdot\chi_{_{([3,1],[3,1])}}+ \chi_{_{([2,2],[2,2])}}
 + \chi_{_{([4],[3,1])}} + \chi_{_{([3,1],[4])}}
 + \chi_{_{([4],[2,2])}}+ \chi_{_{([2,2],[4])}}
+ \chi_{_{([3,1],[2,2])}}+ \chi_{_{([2,2],[3,1])}}
 + \nn \\
+ \chi_{_{([4,1],[4,1])}}+ \chi_{_{([3,2],[3,2])}}
+ \chi_{_{([4,1],[3,2])}}+ \chi_{_{([3,2],[4,1])}}
+ \chi_{_{([4,2],[4,2])}}
\nn \\ \nn \\
\chi_{_{[5,2]}}\cdot\overline{\chi_{_{[5,2]}}} = 1+ 2\cdot\chi_{_{([1],[1])}}
+ 3\cdot \chi_{_{([2],[2])}} + \chi_{_{([1,1],[1,1])}}
+ \chi_{_{([2],[1,1])}}+ \chi_{_{([1,1],[2])}}
 + \nn\\
 +3\cdot \chi_{_{([3],[3])}}+2\cdot\chi_{_{([2,1],[2,1])}}
 + 2\cdot\chi_{_{([3],[2,1])}}+ 2\cdot\chi_{_{([2,1],[3])}}
  + \nn \\ \!\!\!\!\!\!\!\!\!\!\!\!\!\!\!\!\!\!\!\!\!\!\!\!\!\!\!\!\!\!
 + 2\cdot\chi_{_{([4],[4])}} + 2\cdot\chi_{_{([3,1],[3,1])}}+ \chi_{_{([2,2],[2,2])}}
 + 2\cdot\chi_{_{([4],[3,1])}} + 2\cdot\chi_{_{([3,1],[4])}}
 + \chi_{_{([4],[2,2])}}+ \chi_{_{([2,2],[4])}}
+ \chi_{_{([3,1],[2,2])}}+ \chi_{_{([2,2],[3,1])}}
 + \nn \\
  + \chi_{_{([5],[5])}} + 2\cdot\chi_{_{([4,1],[4,1])}}+ \chi_{_{([3,2],[3,2])}}
 + \chi_{_{([5],[4,1])}} + \chi_{_{([4,1],[5])}}
 + \chi_{_{([5],[3,2])}}+ \chi_{_{([3,2],[5])}}
+ \chi_{_{([4,1],[3,2])}}+ \chi_{_{([3,2],[4,1])}}
 + \nn \\
+ \chi_{_{([5,1],[5,1])}}+ \chi_{_{([4,2],[4,2])}}
+ \chi_{_{([5,1],[4,2])}}+ \chi_{_{([4,2],[5,1])}}
+ \chi_{_{([5,2],[5,2])}}
\nn \\ \nn \\
\chi_{_{[6,2]}}\cdot\overline{\chi_{_{[6,2]}}} = 1+ 2\cdot\chi_{_{([1],[1])}}
+ 3\cdot \chi_{_{([2],[2])}} + \chi_{_{([1,1],[1,1])}}
+ \chi_{_{([2],[1,1])}}+ \chi_{_{([1,1],[2])}}
 + \nn\\
 +3\cdot \chi_{_{([3],[3])}}+2\cdot\chi_{_{([2,1],[2,1])}}
 + 2\cdot\chi_{_{([3],[2,1])}}+ 2\cdot\chi_{_{([2,1],[3])}}
  + \nn \\ \!\!\!\!\!\!\!\!\!\!\!\!\!\!\!\!\!\!\!\!\!\!\!\!\!\!\!\!\!\!
 + 3\cdot\chi_{_{([4],[4])}} + 2\cdot\chi_{_{([3,1],[3,1])}}+ \chi_{_{([2,2],[2,2])}}
 + 2\cdot\chi_{_{([4],[3,1])}} + 2\cdot\chi_{_{([3,1],[4])}}
 + \chi_{_{([4],[2,2])}}+ \chi_{_{([2,2],[4])}}
+ \chi_{_{([3,1],[2,2])}}+ \chi_{_{([2,2],[3,1])}}
 + \nn \\ \!\!\!\!\!\!\!\!\!\!\!\!\!\!\!\!\!\!\!\!\!\!\!\!\!\!\!\!\!\!
  + 2\cdot\chi_{_{([5],[5])}} + 2\cdot\chi_{_{([4,1],[4,1])}}+ \chi_{_{([3,2],[3,2])}}
 + 2\cdot\chi_{_{([5],[4,1])}} + 2\cdot\chi_{_{([4,1],[5])}}
 + \chi_{_{([5],[3,2])}}+ \chi_{_{([3,2],[5])}}
+ \chi_{_{([4,1],[3,2])}}+ \chi_{_{([3,2],[4,1])}}
 + \nn \\
  + \chi_{_{([6],[6])}} + 2\cdot\chi_{_{([5,1],[5,1])}}+ \chi_{_{([4,2],[4,2])}}
 + \chi_{_{([6],[5,1])}} + \chi_{_{([5,1],[6])}}
 + \chi_{_{([6],[4,2])}}+ \chi_{_{([4,2],[6])}}
+ \chi_{_{([5,1],[4,2])}}+ \chi_{_{([4,2],[5,1])}}
 + \nn \\
+ \chi_{_{([6,1],[6,1])}}+ \chi_{_{([5,2],[5,2])}}
+ \chi_{_{([6,1],[5,2])}}+ \chi_{_{([5,2],[6,1])}}
+ \chi_{_{([6,2],[6,2])}}
\nn
\ee

}

\item{The starting points for two other series:

{\footnotesize
\be
\phantom.\!
[4,3]\otimes\overline{[4,3]} = ([4,3],[4,3]) + \ \ \ \ \
([4,2],[4,2])+([3,3],[3,3])+([4,2],[3,3])+([3,3],[4,2]) + \nn \\
+([4,1],[4,1])+2\cdot([3,2],[3,2]) + ([4,1],[3,2])+([3,2],[4,1]) + \nn \\
\!\!\!\!\!\!\!\!\!\!\!\!\!\!\!\!\!\!\!\!\!\!\!
+([4],[4])+2\cdot([3,1],[3,1]) + ([2,2],[2,2]) + ([4],[3,1])+([3,1],[4]) +
0\cdot \Big(([4],[2,2])+([2,2],[4])\Big) + ([3,1],[2,2])+([2,2],[3,1]) + \nn \\
+ 2\cdot([3],[3])+2([2,1],[2,1]) + ([3],[2,1])+([2,1],[3]) + \nn \\
+ 2\cdot([2],[2]) + ([1,1],[1,1]) + ([2],[1,1]) + ([1,1],[2]) \ \
+ 2([1],[1]) + (\emptyset,\emptyset) \ \ \ \ \ \
\nn \ee
}

\noindent and

{\footnotesize
\be
\phantom.\!
[3,2,1]\otimes\overline{[3,2,1]} = ([3,2,1],[3,2,1]) + ([3,2],[3,2])+([3,1,1],[3,1,1]) + \ \ \ \ \
\nn \\
\!\!\!\!\!\!\!\!\!\!\!\!\!\!\!\!\!\!\!\!\!\!\!\!\!\!\!\!\!\!\!\!\!\!\!\!\!\!
+([2,2,1],[2,2,1]) + ([3,2],[3,1,1])+([3,1,1],[3,2])
+([3,2],[2,2,1])+([2,2,1],[3,2]) + ([3,1,1],[2,2,1])+([2,2,1],[3,1,1]) +
\nn \\
\!\!\!\!\!\!\!\!\!\!\!\!\!\!\!\!\!\!\!\!\!\!\!\!\!\!\!\!\!\!\!\!\!\!\!\!\!\!
+ 2\cdot\Big(([3,1],[3,1])+  ([2,2],[2,2])+([2,1,1],[2,1,1]) + ([3,1],[2,2])+([2,2],[3,1])
+([3,1],[2,1,1])+([2,1,1],[3,1]) + ([2,2],[2,1,1])+([2,1,1],[2,2])\Big) +
\nn \\
\!\!\!\!\!\!\!\!\!\!\!\!\!\!\!\!\!\!\!\!\!\!\!\!\!\!\!\!\!\!\!\!\!
+ ([3],[3])+6\cdot ([2,1],[2,1])+([1,1,1],[1,1,1]) + 2\cdot ([3],[2,1])+2\cdot([2,1],[3])
+([3],[1,1,1])+([1,1,1],[3]) + 2\cdot([2,1],[1,1,1])+2\cdot([1,1,1],[2,1]) +
\nn \\
\!\!\!\!\!\!\!\!\!\!\!\!\!\!\!\!\!\!\!\!\!\!\!\!\!\!\!\!\!\!\!\!\!
+ 3\cdot\Big(([2],[2])+([1,1],[1,1])+([2],[1,1]) + ([1,1],[2])\Big)
+ 3\cdot([1],[1]) + (\emptyset,\emptyset) \ \ \ \ \ \
\nn
\ee
}
}

\end{itemize}

\noindent
With some abuse of terminology we refer to {\it these} formulas by saying that
representation products are decomposed in this way --
this simplifies the formulations,
but, as we explained in this section, this is not exactly true:
representations contain contributions of diagrams with extra line length,
and accurate statements are about the Schur functions at the $SL_N$ Miwa locus.

\section{Conjecture about the summation domain in differential expansion}

Now, after we got some impression of what the product of representation and its conjugate,
$R\otimes \bar R$ can look like, it is time to formulate our expectation about the
general structure of differential expansion.
The {\bf conjecture} is that HOMFLY and other knot polynomials
{\it always} decompose in exactly this space:
\be
\boxed{
H_R^{\cal K} = \sum_{X\in R\otimes \bar R} Z_R^X F_X^{\cal K}
}
\label{diffexpan}
\ee
with, roughly, the same set of $Z_R^X$ for all knots ${\cal K}$.
We make this claim despite it is anyhow justified only for ${\cal K}$ which
are twist knots, where one indeed deals with the two upcoming braids
and needs to now the decomposition of $R\otimes \bar R$.
For all other knots natural decompositions are very different --
for example, $R^{\otimes m}$ for an $m$-braid knot.
Still, if one believes in universality of differential expansion,
i.e. that it looks the same for all knots, then one can use the formulas
for the twist family as a prototype.
Surprisingly or not, such decompositions indeed exist in many examples,
with the only correction that $Z$-factors actually depend on the {\it defect},
see \cite{Konodef} -- and are exactly the same as for twist knots
only in the case of non-positive defects (zero and minus one),
while get somehow truncated for positive defects, still remaining the same
for all knots with the given defect.
Evidence in support of this mysterious conjecture will be presented elsewhere
\cite{BM}.

\section{Pentad of matrices in non-rectangular case, examples}

Finally we return from the twist-knot polynomials and relatively simple
reduced matrices ${\cal B}^{\rm red}$ to description of the full-fledged
Racah matrices $\bar S$ and $\bar S^{\rm rel}$ and some of their pentad components.
First we study the case of $R=[2,1]$, where the full $\bar S$ is explicitly known,
and then provide a new conjecture for $R=[3,1]$.

\subsection{$R=[2,1]$}

For $R=[2,1]$ we get from explicitly known $\bar S$ of   \cite{Gu} and \cite{M16}:

{
\be
\frac{1}{d_{[2,1]}}\cdot \sum_{\mu,\nu=1}^{10}
(\bar T^{-2}\bar S\bar T^{-2})_{\mu\nu}\,x_\mu x_\nu
\ \ \ = \ \ \ x_1^2
+ \frac{\frac{\{q^3\}\{q\}}{ A^2D_1D_{-1}}}{-A^2}\,\left(\frac{x_4-x_5}{2}\right)^2
+ \frac{\frac{[3]}{[2]^2}D_0^2}{-A^2}\cdot
\left(\frac{A}{D_0}\,x_1 - \frac{\{q\}}{AD_0\sqrt{D_1D_{-1}}}\,x_3\right)^2
+  \nn \\  \!\!\!\!\!\!\!\!
+ \frac{\frac{[3]^2}{[2]^2} D_2D_{-2}}{-A^2}\cdot
\left(\frac{A}{D_0}\,x_1-\frac{\{q\}}{AD_0\sqrt{D_1D_{-1}}}\,
\frac{[2]^2x_2-x_3}{[3]}
+   \frac{\{q^2\}^2(A+A^{-1})}{[3]AD_0\sqrt{D_2D_1D_{-1}D_{-2}}}\,\frac{x_4-x_5}{2}
\right)^2
+ \nn \\ \!\!\!\!\!\!\!\!\!\!\!\!
+ \frac{\frac{[3]}{[2]}D_2D_0D_{-2}D_{-3}}{q^{-2}A^4}\cdot
\left( \frac{q^{-1}A^2}{D_0D_{-1}}\,x_1 - \frac{q^{-1}\{q^2\}}{D_0D_{-2}\sqrt{D_1D_{-1}}}\,x_2
- \frac{q^{-1}\{q \}}{ D_0\sqrt{D_2D_1D_{-1}D_{-2}}}\,\frac{x_4-x_5}{2}
+ \frac{q^3\{q\}\{q^2\}}{A^2D_0D_{-1}D_{-2}\sqrt{D_1D_{-3}}}\,x_6\right)^2
+ \nn \\
+ \frac{\frac{[3]}{[2]}D_3D_2D_0D_{-2}}{q^{2}A^4}\cdot
\left( \frac{qA^2}{D_0D_{1}}\,x_1 - \frac{q\{q^2\}}{D_0D_{2}\sqrt{D_1D_{-1}}}\,x_2
+ \frac{q\{q \}}{ D_0\sqrt{D_2D_1D_{-1}D_{-2}}}\,\frac{x_4-x_5}{2}
+ \frac{q^{-3}\{q\}\{q^2\}}{A^2D_2D_{1}D_{0}\sqrt{D_3D_{-1}}}\,x_7\right)^2
+ \nn \\
+ \frac{D_3D_2D_1D_{-1}D_{-2}D_{-3}}{-A^6}\cdot
\left(\frac{A^3}{D_1D_0D_{-1}}\,x_1 - \frac{\{q^3\}A}{D_2D_0D_{_2}\sqrt{D_1D_{-1}}}\,x_2
+ \frac{q^4\{q\}\{q^3\}}{AD_2D_0D_{-1}D_{-2}\sqrt{D_1D_{-3}}}\,x_6
+ \ \ \ \ \ \ \ \ \ \ \ \right. \nn \\ \left.
+ \frac{q^{-4}\{q\}\{q^3\}}{AD_2D_1D_0D_{-2}\sqrt{D_3D_{-1}}}\,x_7
- \frac{\{q^3\}\{q\}^2}{A^3D_2D_1D_0D_{-1}D_{-2}\sqrt{D_3D_{-3}}}\,x_8\right)^2
+ \nn \\ \!\!\!\!\!\!\!\!\!\!\!\!\!\!\!\!
+ \frac{-\{q^3\}^2\{q\}^2D_2D_{-2}}{-A^6}\left(
\frac{A^3}{D_1D_0D_{-1}}\,x_1 + \frac{A}{\{q^3\}D_0\sqrt{D_1D_{-1}}}\Big(
\left(1-\frac{[4][3]}{[2]}\frac{\{q\}^2}{D_2D_{-2}}\right)\,x_2-x_3\Big)
-\frac{A(A+A^{-1})}{[3]D_0\sqrt{D_2D_1D_{-1}D_{-2}}}\,\frac{x_4-x_5}{2}
-\ \ \ \ \right.\nn \\ \left.
- \frac{q^4}{ [2]AD_0D_{-1}D_{-2}}\sqrt{\frac{D_{-3}}{D_1}}\,x_6
- \frac{q^{-4}}{[2]AD_2D_1D_0}\sqrt{\frac{D_3}{D_{-1}}}\,x_7+
\frac{x_9+x_{10}}{[2]AD_0\sqrt{D_2D_1D_{-1}D_{-2}}}\right)^2
+ \boxed{\frac{d_{[2,1]}^{-1}}{-A^6}\,(x_4+x_5)(x_9-x_{10})}
\nn
\ee
}

\bigskip

\noindent
which we can now compare with (\ref{decobSquadr}) to get the fully-factorized $Z$-factors
and

\be
{\cal E} = U\bar S = d_{[2,1]}\cdot \ \ \ \ \ \ \ \ \ \ \ \ \ \nn
\ee
{\footnotesize
\be
\!\!\!\!\!\!\!\!\!\!\!\!\!\!\!\!\!\!\!\!\!\!\!\!\!\!\!\!\!\!\!\!\!\!\!\!\!\!\!\!\!\!\!
\left(\begin{array}{c|c|cccccc|cc}
1\\
\hline
0&1 \\
\hline
\frac{A}{D_0} &0& -\frac{\{q\}}{AD_0\sqrt{D_1D_{-1}}}
\\
\frac{A}{D_0}  &
*
& \frac{\{q\}}{[3]AD_0\sqrt{D_1D_{-1}}}
& -\frac{[2]^2\{q\}}{[3]AD_0\sqrt{D_1D_{-1}}}
\\
\frac{q^{-1}A^2}{D_0D_{-1}}  &
*
& 0
& \frac{q^{-1}\{q^2\}}{D_0D_{-2}\sqrt{D_1D_{-1}}}
& \frac{q^3\{q\}\{q^2\}}{A^2D_0D_{-1}D_{-2}\sqrt{D_1D_{-3}}}
\\
\frac{q A^2}{D_0D_{1}}  &
*
& 0
& \frac{q \{q^2\}}{D_0D_{2}\sqrt{D_1D_{-1}}}
&0& \frac{q^{-3}\{q\}\{q^2\}}{A^2D_0D_{1}D_{2}\sqrt{D_3D_{-1}}}
\\
\frac{A^3}{D_1D_0D_{-1}}&0&0
&\frac{\{q^3\}A}{D_2D_0D_{-2}\sqrt{D_1D_{-1}}}&
\frac{q^4\{q\}\{q^3\}}{AD_2D_0D_{-1}D_{-2}\sqrt{D_1D_{-3}}} &
\frac{q^{-4}\{q\}\{q^3\}}{AD_2D_1D_0D_{-2}\sqrt{D_3D_{-1}}} &
-\frac{\{q\}^2\{q^3\}}{A^3D_2D_1D_0D_{-1}D_{-2}\sqrt{D_3D_{-3}}}\!\!\!\!\!\!\!\! &
\\
\frac{A^2}{D_1D_0D_{-1}}  &
*
& -\frac{A}{\{q^3\}D_0\sqrt{D_1D_{-1}}}
&\frac{A\left(1-\frac{[4][3]}{[2]}\frac{\{q\}^2}{D_2D_{-2}}\right)}{\{q^3\}D_0\sqrt{D_1D_{-1}}} &
-\frac{q^4}{[2]AD_0D_{-1}D_{-2}}\sqrt{\frac{D_{-3}}{D_1}} &
\frac{q^{-4}}{[2]AD_2D_1D_0}\sqrt{\frac{D_3}{D_{-1}}} & \!\!\!\!\!\! 0 &
\!\!\!\!\!\!\!\!\!\!\!\!\!\!\!\! \!\!\!\!\!\!
\frac{1}{\sqrt{2}[2]AD_0\sqrt{D_2D_1D_{-1}D_{-2}}} \\
\hline
\ldots &
\end{array}\right)
\nn
\ee
}

\bigskip

\noindent
Here the obvious reordering was made w.r.t. conventions of \cite{Gu,M16}
to obtain a triangular matrix:

the first line/column corresponds to $x_1$,

the second one --  to the combination $x_4-x_5$,

the third one -- to $x_3$,

the forth one -- to $x_2$,

the fifth one -- to $x_6$,

the sixth one -- to $x_7$,

the seventh one -- to $x_8$,

the eighth one -- to the combination $x_{9}+x_{10}$.

We omitted the last two lines, associated to $x_4+x_5$
and $x_9-x_{10}$, where we have an "irrelevant" {\it non-triangular} $2\times 2$ block,
decoupled from the rest of the matrix.

Suppressed (to save the space) is also the second column, which is actually
\be
\left(\begin{array}{c}
0\\  1 \\ 0\\ * \\ * \\ * \\ 0 \\ *
\end{array}\right) =
\left(\begin{array}{c}
0\\ 1\\ 0 \\ \frac{\{q^2\}^2(A+A^{-1})}{\sqrt{2}[3]AD_0\sqrt{D_2D_1D_{-1}D_{-2}}} \\
-\frac{q^{-1}\{q^3\}}{\sqrt{2}[3]D_0\sqrt{D_2D_1D_{-1}D_{-2}}} \\
-\frac{q \{q^3\}}{\sqrt{2}[3]D_0\sqrt{D_2D_1D_{-1}D_{-2}}} \\
0 \\ \frac{A(A+A^{-1})}{\sqrt{2}[3]D_0\sqrt{D_2D_1D_{-1}D_{-2}}}
\end{array}\right)
\ee

\bigskip

It is easy to observe that the sum of the entries in the third and the forth columns
reproduce the answers in the second column of (\ref{E21}),
describing reduced ${\cal E}^{\rm red}$ in the $6\times 6$ formalism.
More accurately,
reduction from $10\times 10$ to $6\times 6$ matrices
with omitted lines and columns $\#\ 2,3,9,10$
and the entries of columns 3 and 4 added,
is provided by multiplication from
the left and from the right  by respectively
\be
P_L=\left(\begin{array}{cccccccccc}  {\bf 1} &0&0&0&0&0&0&0&0&0\\
0&0&0& {\bf 1} &0&0&0&0&0&0\\
0&0&0&0& {\bf 1} &0&0&0&0&0\\
0&0&0&0&0& {\bf 1} &0&0&0&0\\
0&0&0&0&0&0& {\bf 1} &0&0&0 \\
0&0&0&0&0&0&0& {\bf 1} & 0 & 0
\end{array}\right)
\ \ \ \ \ \ \ {\rm and} \ \ \ \ \ \ \ \ \
P_R=\left(\begin{array}{cccccc} {\bf 1}&0&0&0&0&0 \\  0& 0&0&0&0&0 \\
0&{\bf 1}&0&0&0&0 \\ 0&{\bf 1}&0&0&0&0\\
0&0& {\bf 1}&0&0&0\\0&0&0& {\bf 1}&0&0\\0&0&0&0& {\bf 1}&0\\0&0&0&0&0& {\bf 1}\\
0&0&0&0&0&0 \\ 0&0&0&0&0&0
\end{array}\right)
\ee
The product $P_LP_R$ equals the $6\times 6$ unit matrix,
but the product $P_RP_L$ is not quite a projection matrix in $10$-dimensional space,
however, it acts as unit matrix in the $6$-dimensional one.

\bigskip

For example,
$$P_L (U\bar S) P_R = {\cal E}^{\rm red}_{[2,1]} =
$$

\bigskip

\centerline{{\footnotesize
$
\ \ = d_{[2,1]}\cdot\left(\begin{array}{cccccc}
1&0&0&0&0&0 \\ \frac{A}{D_0}& -\frac{\{q\}}{AD_0\sqrt{D_1D_{-1}}} & 0&0&0&0\\
\frac{A^2}{qD_0D_{-1}} & -\frac{\{q^2\}}{qD_0D_{-2}\sqrt{D_1D_{-1}}}&
\frac{q^3\{q\}\{q^2\}}{A^2D_0D_{-1}D_{-2} \sqrt{D_1D_{-3}}}&0&0&0 \\
\frac{qA^2}{D_0D_1 }& -\frac{q\{q^2\}}{D_0D_2\sqrt{D_1D_{-1}}} & 0 &
\frac{\{q\}\{q^2\}}{q^3A^2D_0D_1D_2\sqrt{D_3D_{-1}}} & 0 & 0 \\
\frac{A^3}{ D_1D_0D_{-1}} & -\frac{A\{q^3\}}{D_2D_0D_{-2}\sqrt{D_1D_{-1}}} &
\frac{q^4\{q\}\{q^3\}}{AD_2D_0D_{-1}D_{-2}\sqrt{D_1D_{-3}}} &
\frac{\{q\}\{q^3\}}{q^4AD_2D_1D_0D_{-2}\sqrt{D_3D_{-1}}} &
-\frac{\{q\}^2\{q^3\}}{A^3D_2D_1D_0D_{-1}D_{-2}\sqrt{D_3D_{-3}}}& 0 \\
\frac{A^3}{D_1D_0D_{-1}}&
-\frac{A[4]\{q\}}{[2]D_2D_0D_{-2}\sqrt{D_1D_{-1}}}&
-\frac{q^4 }{[2]AD_0D_{-1}D_{-2}}\sqrt{\frac{D_{-3}}{D_1}} &
-\frac{1}{q^4[2]AD_2D_1D_0}\sqrt{\frac{D_3}{D_{-1}}}&
0 & \!\!\!\!\!\!\!\!\!\!\!\!\!\!\!\!
\frac{\sqrt{2}}{[2]AD_0\sqrt{D_2D_1D_{-1}D_{-2}}}
\end{array}\right)
$
}}

\bigskip

\noindent
which is exactly the reduced (\ref{E21}).
We see that the first five lines are in close correspondence with the table
for $R=[2,2]$, and that the $Z$-factors are properly reproduced.
Moreover, the non-factorized $Z$-factor for $\lambda=[1]$ (originally suggested in \cite{Ano21})
is naturally decomposed into two contributions:
\be
Z^{[2,1]}_{[1]} \ = \  D_3D_{-3}+D_2D_0+D_0D_{-2} \ = \
\frac{[3]}{[2]^2}D_0^2 + \frac{[3]^2}{[2]^2} D_2D_{-2}
\ee
From our experience with the case of rectangular $R=[r^s]$ we could expect that this
is decomposed as
\be
\frac{1}{d_{[2,1]}}\cdot \sum_{\mu,\nu=1}^{10}
(\bar T^{-2}\bar S\bar T^{-2})_{\mu\nu}\,x_\mu x_\nu
\ \ \ \stackrel{?}{=} \ \ \
\sum_{\lambda=1}^{10} \frac{Z_\lambda}{\Lambda_\lambda'}\cdot
\left(\sum_{\mu=1}^{\lambda} \frac{(U\bar S)_{\lambda\mu} x_\mu}{d_{[2,1]}}\right)^2
\label{decoSTS21}
\ee
but this is not fully true:
the term in the box is not of this form.
What happens is that the $10\times 10$ matrix $S$ has a block form,
with non-trivial $8\times 8$ block and a trivial, but non-unit $2\times 2$:
\be
\bar S = \bar S_8 \otimes \left(\begin{array}{cc} 0&1\\1& 0 \end{array}\right)
\ee
The same is true for $\bar T^{-2}\bar S \bar T^{-2}$.
The point is that the matrix $\left(\begin{array}{cc} 0&1\\1& 0 \end{array}\right)$
does not possess Gauss decomposition.
However, since it is a separate block, it does not interfere with the $8\times 8$ sector
and vanishes after projection onto the state $|\emptyset>$, which belongs to the $8$-sector.
This means that we actually expect (\ref{decoSTS21}) to be true in the $8\times 8$ sector,
ortgogonal to $x_4+x_5$ and $x_9-x_{10}$ --
where it is indeed true.
Note that in these coordinates
$\bar T$ is not diagonal, only $\bar T^2$ is.
The relevant fragment in diagonal  $\bar T$, was
  $\left(\begin{array}{cccc} A^2 \\ & -A^2 \\ & & A^4 \\ &&& -A^4 \end{array}\right)$,
  and it is now substituted by
  $\left(\begin{array}{cccc} 0& A^2 \\ A^2&0 \\ & & 0& A^4 \\ &&A^4 & 0 \end{array}\right)$.

\bigskip

The full $10\times 10$ matrix ${\cal B}$ in these coordinates is also of the block form:

{\footnotesize
\be
\!\!\!\!\!\!\!\!\!\!\!\!\!\!\!\!\!\!\!\!\!\!\!
{\cal B}_{[2,1]}= \left(\begin{array}{c||c||c||c|c|c|c|c|c||cc}
&\emptyset& [1]_z & [1]_y & [1]_x & [2] & [1,1] & [2,1] & X_2 &&\\
&&&&&&&& \\
\hline\hline
&&&&&&&& \\
\emptyset & 1 &&&&&&& \\
&&&&&&&& \\
\hline\hline
&&&&&&&& \\
\left[1\right]_z&0& A^2&&& &&&&&\\
&&&&&&&& \\
\hline\hline
&&&&&&&& \\
\left[1\right]_y& -A^2&0&A^2 &&&&&& \\
&&&&&&&& \\
\left[1\right]_x &-A^2&0&0&A^2 &&&&&\\
&&&&&&&& \\
\hline
&&&&&&&& \\
\left[2\right] &q^2A^4
& -q^3A^4[N-2]
& - \frac{q^3A^4}{[2]}
& -\frac{q^3[3]A^4}{[2]}& q^4A^4 &&  &&\\
&&&&&&&& \\
\hline
&&&&&&&& \\
\left[1,1\right] &\frac{A^4}{q^2}
& \frac{A^4[N+2]}{q^3}
& -\frac{A^4}{[2]q^3}
& - \frac{[3]A^4}{[2]q^3} & 0 & \frac{A^4}{q^4} &&&\\
&&&&&&&& \\
\left[2,1\right] & -A^6
& -[3]A^5
& \frac{[3]A^6}{[2]^2}
& \frac{[3]^2A^6}{[2]^2} & -\frac{[3]qA^6}{[2]}& -\frac{[3]A^6}{q[2]}&A^6 && \\
&&&&&&&& \\
\hline
&&&&&&&& \\
X_2 &\overline{-A^6}
& \overline{-[3]A^5}+\frac{A^6[N]}{\{q\}}
&\overline{\frac{[3]A^6}{[2]^2}}- \frac{ A^5[N]}{[2]^2\{q\}}
&\overline{\frac{[3]^2A^6}{[2]^2}}  + \frac{ A^5[N]}{[2]^2\{q\}}
&\underbrace{-\frac{qA^5[N+3]}{[2] }}_{\overline{-\frac{[3]qA^6}{[2]}} -\frac{A^5[N]}{q^2[2]}  }
&\underbrace{\frac{A^5[N-3]}{q[2] }}_{\overline{-\frac{[3]qA^6}{[2]}}+\frac{q^2A^5[N]}{[2]}   }
&0&A^4 &\\
&&&&&&&& \\
\hline
&&&&&&&& \\
&0&0&0&0&0&0&0&0&A^4&0\\
&0&0&0&0&0&0&0&0&0&A^2
\end{array}\right)
\nn
\ee
}

\bigskip

\noindent
and it also reproduces reduced ${\cal B}^{\rm red}$ from (\ref{Bred21})
after rejecting the third, ninth and tenth lines/columns,
and  adding the entries in the second and forth columns and lines.
Overlined items in the line $X_2$ coincide with the entries of the previous line $[2,1]$ --
this is one of the structures, convenient in the search for generalizations
to higher representations $R$.

Also, to see the universality it is useful to compare this ${\cal B}_{[2,1]}$ to
a simpler matrix in the case of $R=[1,1]$ without multiplicities:

{\footnotesize
\be
{\cal B}_{[1,1]} =
\left(\begin{array}{c||ccc}
&\emptyset & [1] & [1,1] \\
&&&\\
\hline\hline
&&&\\
\emptyset & 1 &&\\
&&&\\
\left[1\right] & -A^2 & A^2 & \\
&&&\\
\left[1,1]\right] & \frac{A^4}{q^2} & -\frac{[2]A^4}{q^3} & \frac{A^4}{q^4} \\
&&&\\
\end{array}\right)
\ee
}

\subsection{$R=[3,1]$
\label{31}}

In this case
\be
\phantom.
[3,1]\otimes\overline{[3,1]} = (\emptyset,\emptyset) +
4\cdot([1],[1]) + 4\cdot ([2],[2]) + ([1,1],[1,1]) + \underline{([2],[1,1])+([1,1],[2])}
+ \nn \\
+ ([3],[3]) + ([2,1],[2,1])+ \underline{([3],[2,1])+([2,1],[3])} + ([3,1],[3,1])
\ee
contains two representations with quadruple multiplicities and two pairs of non-diagonal
composites (underlined), this the full matrices of the pentad will be
$17\times 17$.
However, the two $2\times 2$ blocks decouple, so that {\it relevant} matrices
will be $13\times 13$, and if one is interested in twist-knot polynomials only,
${\cal B}$ can be further {\it reduced} to $9\times 9$.
In this reduced case we will get two non-factorized $Z$-factors, which are the linear
combinations of factorized ones from the {\it relevant} level of $13\times 13$.

Since compact notation is not yet invented to described the complicated formulas
in the $R=[r,1]$ case, we present at the end of this paper 
just the simplest of the {\it relevant} matrices:
${\cal B}_{[3,1]}^{\rm rel}$.
Reduced $9\times 9$ matrix ${\cal B}^{\rm red}_{[3,1]}$,
needed for the twist-knot calculus and explicitly provided in \cite{M19nr} and 
in s.\ref{redB} above, is obtained from it by omission of the
 lines/columns $[1]_z$ and $[2]_z$
and by summing up the entries of the columns $[1]_x+[1]_y$ and $[2]_x+[2]_y$.
To obtain Racah matrices $\bar S$ and $S$, needed for arborescent calculus and
originally found in \cite{MnonrectRacah} one should follow the sequence of steps,
described in s.\ref{Eform}:
solve (\ref{Eeigen}) to obtain ${\cal E}$,
then normalize it properly to satisfy (\ref{srules}) and
use it to define $\bar S$ as a quadratic form (\ref{decobSquadr}),
and finally find $S$ from its diagonalization (\ref{SfrobS}). 
All these are straightforward linear-algebra operations,
moreover, they remain just the same for all other representations $R$ --
in variance of artistic result of \cite{MnonrectRacah}, which is very difficult
to generalize.
Thus what we now need for other non-rectangular $R$ are the bigger pieces
of the universal ${\cal B}^{\rm rel}$, 
which would include all representations from $R\otimes \bar R$.

\section{Conclusion}

To conclude, we made a new small step in the difficult study of differential expansions
for knot polynomials.
They are now very well described (though not fully understood) for the intersection
of two domains: of twist knots and of rectangular representations.
To be precise,
summation set ${\cal R}_R$,  $Z$-factors and $F$-coefficients in
\be
H_R^{\cal K}(A,q) = \sum_{X\in {\cal R}_R} Z_R^X(A,q) F_X^{\cal K}(A,q)
\ee
are explicitly known for ${\cal K}\subset {\rm twist}\ \&\ {\rm double\ braid\ knots}$
and for $R=[r^s]$.
The two obvious directions to generalize are to arbitrary ${\cal R}_R$ and to arbitrary knots.
For the first task we suggest to simultaneously deform the entire pentad structure \cite{M19pentad},
which complements Racah matrices $\bar S$ and $S$ by $U$ and, most important, by
two {\it triangular} and {\it universal} ($R$-independent) matrices ${\cal B}$ and ${\cal E}$.
As a new step in this direction we lift the previously known result for the simplest
non-rectangular $R=[2,1]$ to $R=[3,1]$.
Concerning generalization to other knots, we begin with conjecturing in (\ref{diffexpan})
the {\it universality} (${\cal K}$-independence) of the domain ${\cal R}_R$, to begin with:
${\cal R}_R \ \stackrel{?}{=}\ R\otimes\bar R$.
Now we have all the notions and means to proceed for technical calculations,
which are going to cover more general representations and more general knows.

\section*{Appendix}

In this appendix we present the fragment of ${\cal B}^{rel}$, including all the representations
from $[3,1]\times\overline{[3,1]}$. 
Its role and applications are described in s.\ref{31}.

\begin{sidewaystable}
{\tiny
$$
\left(\begin{array}{c||c||c|c||c|c|c|c|c| }
&\emptyset& [1]_z & [2]_z & [1]_y & [2]_y & [1]_x & [2]_x &
\\
&&&&&&&  \\
\hline\hline
&&&&&&&  \\
\emptyset & 1 &&&&&&  \\
&&&&&&&  \\
\hline\hline
&&&&&&&  \\
\left[1\right]_z&0& A^2&&&&&  \\
&&&&&&&  \\
\left[2\right]_z&0& -q^2A^3  & q^4A^4  &&&&  \\
&&&&&&&  \\
\hline\hline
&&&&&&&  \\
\left[1\right]_y& -A^2&0& 0&A^2 &&&  \\
&&&&&&&  \\
\left[2\right]_y& q^2A^4&\underline{-q^3A^4[N-2]} & 0
&-[2]q^3A^4 & q^4A^4  &&  \\
&&&&&&&  \\
\left[1\right]_x &-A^2&0&0&0&0&A^2 &  \\
&&&&&&&  \\
\left[2\right]_x &q^2A^4&\underline{-q^3A^4[N-2]}&0
&\underline{-\frac{q^3A^4}{[2]}}&0&\underline{-\frac{q^3[3]A^4}{[2]}} &q^4A^4&  \\
&&&&&&&  \\
\hline
&&&&&&&  \\
\left[3\right] &-q^6A^6& q^8A^6[2][N-2] & q^{11}A^7[N-2]
& q^8A^6 & -\frac{q^{10}A^6}{[3]} & \frac{[4]q^8A^6}{[2]}
& -\frac{[4][2]q^{10}A^6}{[3]}&  \\
&&&&&&&  \\
\hline
&&&&&&&  \\
\left[1,1\right] &\frac{A^4}{q^2}
& \frac{[2]A^4[N+3]}{q^3}
& 0 & -\frac{[2]A^4}{q^3[3]} & 0 & -\frac{[4]A^4}{q^3[3]} & 0 &  \\
&&&&&&&  \\
\left[2,1\right] & -A^6
&   {-[3]A^5} - \frac{A^6[N+4]}{q}
& -\frac{q^{2}A^7[N+4]}{[2] }
& A^6 & -\frac{qA^6}{[3][2]} & \frac{[4]A^6}{[2]}
&-\frac{[4]qA^6}{[3]} &  \\
&&&&&&&  \\
\left[3,1\right] &q^4A^8
& q^4[4][2]A^7
& q^{7}[4]A^8
& -\frac{[4]q^5A^8}{[3]} & \frac{[4]q^7A^8}{[3]^2} & -\frac{[4]^2q^5A^8}{[3][2]}
& \frac{[4]^2[2]q^7A^8}{[3]^2} &  \\
&&&&&&&  \\
\hline
&&&&&&&  \\
X_2 & {-A^6}
&   {-[3]A^5} - \frac{A^6[N+4]}{q}+{\frac{A^6[N]}{\{q\}}}
&  -\frac{q^{2}A^6[N+4][N+1]}{[2]}
&  {A^6}-\frac{A^5[N]}{q[3][2]\{q\}}
&   -\frac{qA^5[N+1]}{[3][2]}
&  {\frac{[4]A^6}{[2]}} + \frac{A^5[N]}{q[3][2]\{q\}}
& -\frac{qA^5[N+4]}{[3]}  &   \\
&&&&&&&  \\ &&&&&&&  \\
X_3 &  {q^4A^8} &  {q^4[4][2]A^7} -\frac{q^5A^8[N+1]}{\{q\}}
&  {q^7[4]A^8}-\frac{q^6A^8[N+2][N+1]}{[2]} - \frac{q^3A^7[N+1]}{\{q\}}
&  {-\frac{[4]q^5A^8}{[3]}}+\frac{q^4A^7[N+1]}{[3][2]]\{q\}}
&  {\frac{[4]q^7A^8}{[3]^2}}-\frac{[4]q^4A^7[N+1]}{[3]^2[2]^2\{q\}}
&  {-\frac{[4]^2q^5A^8}{[3][2]}} -\frac{q^4A^7[N+1]}{[3][2]]\{q\}}
&   {\frac{[4]^2[2]q^7A^8}{[3]^2}}+ \frac{q^3A^7[N+1]}{[2]}
+\frac{[4]q^4A^7[N+1]}{[3]^2[2]^2\{q\}}
&   \\
&&&&&&&  \\
\hline\hline
&&&&&&&  \\
&\ldots&&&&&&  \\
\end{array}\right.
$$
}
{\tiny
$$
\left.\begin{array}{c||c|c|c|c|c|c|c||c}
&\ldots&[3]&[1,1]&[2,1]&[3,1]&X_2&X_3&\ldots \\
&&&&&&&& \\
&\ldots&&&&&&& \\
&&&&&&&& \\
\hline
&&&&&&&& \\
\left[3\right] &\ldots
&q^{12}A^6&&&&& \\
&&&&&&&& \\
\hline
&&&&&&&& \\
\left[1,1\right] &\ldots
& 0 &\frac{A^4}{q^4}&&&&\\
&&&&&&&& \\
\left[2,1\right] &\ldots
& 0 & -\frac{[3]A^6}{q[2]} & A^6&&&& \\
&&&&&&&& \\
\left[3,1\right] &\ldots
& -\frac{[4]q^9A^8}{[3]} & \frac{[4]q^4A^8}{[2]}
& -\frac{[4][2]q^6A^8}{[3]}  & q^8A^8 &&&\\
&&&&&&&& \\
\hline
&&&&&&&& \\
X_2  &\ldots
& 0
& -\frac{A^5[N-3]}{q[2]}
& 0 & 0 & A^4 && \\
&&&&&&&& \\ &&&&&&&& \\
X_3 &\ldots
&  -\frac{q^8A^7[N+5]}{[3]}
& -\frac{q^3A^7[N-3]}{[2]}
&
\frac{q^5A^7[N-1][N-3]}{[3][N]}
&0&-\frac{q^5A^6[N+1]}{[N]}&q^6A^6\\
&&&&&&&& \\
\hline\hline
&&&&&&&& \\
&\ldots&&&&&&&\ldots \\
\end{array}\right.
$$
}
\end{sidewaystable}

\newpage

\section*{Acknowledgements}

This work was partly supported
by the Russian Science Foundation (Grant No.16-12-10344).


\begin{thebibliography}{12}

\bibitem{DGR} N.M.Dunfield, S.Gukov  and J.Rasmussen, Experimental Math. 15 (2006) 129-159,
math/0505662

\bibitem{IMMMfe} H. Itoyama, A. Mironov, A. Morozov and An. Morozov, JHEP 2012 (2012) 131,
arXiv:1203.5978

\bibitem{evo} A. Mironov, A. Morozov and An. Morozov,
AIP Conf. Proc. 1562 (2013) 123, arXiv:1306.3197 \!\!;
Mod. Phys. Lett. A 29 (2014) 1450183,  arXiv:1408.3076

\bibitem{diffarth} S.Arthamonov, A.Mironov, A.Morozov,
Theor.Math.Phys. 179 (2014) 509-542, arXiv:1306.5682

\bibitem{knotpols}
J.W.Alexander, Trans.Amer.Math.Soc. 30 (2) (1928) 275-306\\
V.F.R.Jones, Invent.Math. 72 (1983) 1 Bull.AMS 12 (1985) 103 Ann.Math. 126 (1987) 335\\
L.Kauffman, Topology 26 (1987) 395\\
P.Freyd, D.Yetter, J.Hoste, W.B.R.Lickorish, K.Millet, A.Ocneanu, Bull. AMS. 12 (1985) 239\\
J.H.Przytycki and K.P.Traczyk, Kobe J Math. 4 (1987) 115-139\\
A.Morozov, Theor.Math.Phys. 187 (2016) 447-454, arXiv:1509.04928

\bibitem{Wit}
E. Witten, Comm.Math.Phys. 121 (1989) 351-399

\bibitem{ind}
R.K. Kaul, T.R. Govindarajan, Nucl.Phys. B380 (1992) 293-336, hep-th/9111063 \\
P. Ramadevi, T.R. Govindarajan, R.K. Kaul, Nucl.Phys. B402 (1993) 548-566, hep-th/9212110; Nucl.Phys.
B422 (1994) 291-306, hep-th/9312215 \\
P. Ramadevi, T. Sarkar, Nucl.Phys. B600 (2001) 487-511, hep-th/0009188

\bibitem{RT}
 E. Guadagnini, M. Martellini, M. Mintchev, Clausthal 1989, Proceedings, Quantum groups, 307-317;
Phys.Lett. B235 (1990) 275 \\
N.Yu. Reshetikhin, V.G. Turaev, Comm.Math.Phys. 127 (1990) 1-26

\bibitem{Racah}
G. Racah, 
Phys.Rev. {\bf 62} (1942) 438-462\\
E.P. Wigner, Manuscript, 1940,  in: {\sl Quantum Theory of Angular Momentum},
pp. 87–133, Acad.Press, 1965;
{\sl Group Theory and Its Application to the Quantum Mechanics of Atomic Spectra},
Acad.Press,   1959\\
L.D. Landau and E.M. Lifshitz, {\sl Quantum Mechanics: Non-Relativistic Theory},
Pergamon Press, 1977 \\
J. Scott Carter, D.E. Flath, M. Saito, {\sl The Classical and Quantum 6j-symbols},
Princeton Univ.Press, 1995 \\
S. Nawata, P. Ramadevi and Zodinmawia, Lett.Math.Phys. {\bf 103} (2013) 1389-1398,
arXiv:1302.5143 \\
A. Mironov, A. Morozov, A. Sleptsov,  	JHEP 07 (2015) 069, arXiv:1412.8432 \\
V.Alekseev, An.Morozov and A.Sleptsov,  arXiv:1909.07601;  arXiv:1912.13325

\bibitem{arbor}
A.Mironov, A.Morozov, An.Morozov, P.Ramadevi, V.K.Singh,
JHEP {\bf 1507} (2015) 109,  arXiv:1504.00371 \\
S.Nawata, P.Ramadevi, V.K.Singh,  arXiv:1504.00364 \\
A.Mironov and A.Morozov, Phys.Lett. B755 (2016) 47-57, arXiv:1511.09077

\bibitem{M16} A.Morozov,  JHEP 1609 (2016) 135, arXiv:1606.06015 v8

\bibitem{KNTZ}
M.Kameyama, S.Nawata, R.Tao, H.D.Zhang,  arXiv:1902.02275

\bibitem{M19} A.Morozov, Phys.Lett. B 793 (2019) 116-125, arXiv:1902.04140

\bibitem{M19nr} A.Morozov, Phys.Lett. B 793 (2019) 464-468, arXiv:1903.00259

\bibitem{M19pentad} A.Morozov,  Eur.Phys.J.Plus (2020),
DOI :10.1140/epjp/s13360-020-00234-w,  arXiv:1906.09971

\bibitem{GSV} S.Gukov, A.Schwarz and C.Vafa, Lett.Math.Phys. 74 (2005) 53-74, arXiv:hep-th/0412243

\bibitem{DMMSS}
M.Aganagic, Sh.Shakirov, arXiv:1105.5117 \!\!;  arXiv:1202.2489 \!\!; arXiv:1210.2733 \\
P.Dunin-Barkowski, A.Mironov, A.Morozov, A.Sleptsov, A.Smirnov, JHEP 03 (2013) 021,
arXiv:1106.4305 \\
I. Cherednik, arXiv:1111.6195

\bibitem{Okshift}
A.Okounkov, G.Olshanksy, Algebra i Analiz 9 (1997) No.2;
Math.Res.Lett. 4 (1997) 69-78, q-alg/9608020

\bibitem{Okmac} A.Okounkov, arXiv:q-alg/9608021

\bibitem{GGS} E.Gorsky, S.Gukov,  M.Stosic,
Fundamenta Mathematicae 243 (2018) 209–299, arXiv:1304.3481

\bibitem{NawOb} S.Nawata and A.Oblomkov,
Contemp. Math. 680 (2016) 137, arXiv:1510.01795

\bibitem{Anokhevo}
A.Anokhina, A.Morozov, JHEP 1804 (2018) 066, arXiv:1802.09383  \\
P.Dunin-Barkowski, A.Popolitov, S.Popolitova, arXiv:1812.00858 \\
A.Anokhina, A.Morozov, A.Popolitov, Eur.Phys.J.C (2019),  arXiv:1904.10277 and {\it to appear}

\bibitem{pretzel}
D.Galakhov, D.Melnikov, A.Mironov, A.Morozov, A.Sleptsov,
 Phys.Lett. B743 (2015) 71, arXiv:1412.2616 \\
A.Mironov, A.Morozov, A.Sleptsov,
JHEP 07 (2015) 069, arXiv:1412.8432 \\
D.Galakhov, D.Melnikov, A.Mironov and A.Morozov,
Nucl.Phys. B 899 (2015) 194-228, arXiv:1502.02621 \\
A.Mironov, A.Morozov, An.Morozov, A.Sleptsov,
JETP Lett. 104 (2016) 56-61, Pisma Zh.Eksp.Teor.Fiz. 104 (2016) 52-57,  arXiv:1605.03098 \\
Sh. Shakirov and A. Sleptsov,
arXiv:1611.03797 \\
S.Arthamonov and Sh.Shakirov,  arXiv:1704.02947

\bibitem{M1605} A.Morozov,  Nucl.Phys. B911 (2016) 582-605, arXiv:1605.09728

\bibitem{KM17fe}  Ya.Kononov and A.Morozov,
Theor.Math.Phys. 193 (2017) 1630-1646,
arXiv:1609.00143

\bibitem{KM17tw} Ya.Kononov and A.Morozov,
Mod.Phys.Lett. A Vol. 31, No. 38 (2016) 1650223,
arXiv:1610.04778

\bibitem{Mnonrect} A.Morozov, Mod.Phys.Lett. A33 No. 12 (2018) 1850062, arXiv:1612.00422

\bibitem{MnonrectRacah}
A.Morozov, Phys.Lett. B 766 (2017) 291-300, arXiv:1701.00359

\bibitem{M333} A.Morozov,  Phys.Lett. B778 (2018) 426-434,  arXiv:1711.09277

\bibitem{arborgauge}
A. Mironov, A. Morozov, An. Morozov, P. Ramadevi, V.K. Singh and A. Sleptsov,
J.Phys. A: Math.Theor. {\bf 50} (2017) 085201, arXiv:1601.04199

\bibitem{Koj}  K.Koike, Adv. Math. 74 (1989) 57

\bibitem{MMhopf}  H.Kanno, Nucl.Phys. B745 (2006) 165-175, hep-th/0602179 \\
 A.Mironov and A.Morozov, JETP Lett. 107 (2018) 728-735, arXiv:1804.10231;
 Nucl.Phys. B 944 (2019) 114641, arXiv:1903.00773 \\
 H.Awata, H.Kanno, A.Mironov and A.Morozov,   Nucl.Phys. B 949 (2019) 114816,  arXiv:1905.00208

\bibitem{Konodef} Ya.Kononov and A.Morozov, JETP Lett. 101 (2015) 831-834, arXiv:1504.07146

\bibitem{BM} L.Bishler et al., {\it to appear}

\bibitem{Gu}  J.Gu and H.Jockers, Commun.Math.Phys. 338 (2015) 393-456, arXiv:1407.5643

\bibitem{Ano21} A.Anokhina, A.Mironov, A.Morozov and An.Morozov,
Nucl.Phys. B 882C (2014) 171-194,  arXiv:1211.6375



\end{thebibliography}
\end{document}